\documentclass[aps,prc,twocolumn,amsmath,floatfix,nofootinbib]{revtex4-2}

\usepackage{amsmath}
\usepackage{bm}
\usepackage{graphicx}
\usepackage[urlcolor=green]{hyperref}
\usepackage[dvipsnames]{xcolor}

\begin{document}

\author{Nicolas Borghini}
\email{borghini@physik.uni-bielefeld.de}
\author{Marc Borrell} \email{marcborrell@physik.uni-bielefeld.de}
\author{Hendrik Roch}
\email{hroch@physik.uni-bielefeld.de}
\affiliation{Fakult\"at f\"ur Physik, Universit\"at Bielefeld, 
	D-33615 Bielefeld, Germany}

\title{Early time behavior of spatial and momentum anisotropies in kinetic theory across different Knudsen numbers}

\begin{abstract}
We investigate the early time development of the anisotropic transverse flow and spatial eccentricities of a fireball with various particle-based transport approaches using a fixed initial condition.
In numerical simulations ranging from the quasi-collisionless case to the hydrodynamic regime, we find that the onset of $v_n$ and of related measures of anisotropic flow can be described with a simple power-law ansatz, with an exponent that depends on the amount of rescatterings in the system. 
In the few-rescatterings regime we perform semi-analytical calculations, based on a systematic expansion in powers of time and the cross section, which can reproduce the numerical findings. 
\end{abstract}

\maketitle

\section{Introduction}
\label{s:intro}

Collisions of heavy nuclei at high energies create a highly dynamical system, which develops some collective behavior over a timescale of order 10\;fm$/c$. 
The emission pattern of particles in the final state appears to be strongly correlated to the initial system geometry determined by the overlap region of the colliding nuclei. 
In particular, initial asymmetries in the geometry are converted into transverse momentum space anisotropies, referred to as anisotropic flow, as the system evolves~\cite{Heinz:2013th,Luzum:2013yya,Bhalerao:2020ulk}.

When two nuclei collide, the transverse geometry of their overlap region is customarily characterized in the transverse plane by spatial eccentricities.
In polar coordinates $(r,\theta)$, these are defined by~\cite{Alver:2010gr,Teaney:2010vd,Gardim:2011xv}
\begin{align}
	\label{eccentricity}
\varepsilon_n^{\bf x} {\rm e}^{{\rm i}n\Phi_n}\equiv 
-\frac{\langle r^n {\rm e}^{{\rm i}n\theta}\rangle_{\bf x}}{\langle r^n\rangle_{\bf x}},
\end{align}
where $\langle\cdots\rangle_{\bf x}$ is an average over the transverse plane --- ${\bf x}$ denotes the transverse position vector ---, weighted with the centered entropy or energy density.\footnote{In the paper, we will consider averages weighted with particle number or energy.}
The angle $\Phi_n$ is the $n$-th participant plane angle, which we shall set equal to 0 in the following.

In turn, momentum space anisotropies are mostly characterized by the Fourier coefficients of the transverse momentum distribution of emitted particles~\cite{Voloshin:1994mz}:
\begin{align}
v_n{\rm e}^{{\rm i}n\Psi_n}\equiv 
  \langle {\rm e}^{{\rm i}n\varphi_\mathbf{p}}\rangle_{\bf p},
\label{vn_def}
\end{align}
where $\varphi_\mathbf{p}$ is the azimuth of a particle transverse momentum ${\bf p}$ and $\Psi_n$ is the $n$-th ``event plane'' angle.
The angular brackets now denote an average over the momentum distribution of particles.

Most of the modern models of the dynamics of the fireball created in high-energy nuclear collisions include a fluid dynamical stage~\cite{Ollitrault:1992bk,Kolb:2003dz,Huovinen:2006jp,Romatschke:2009im,Teaney:2009qa,Romatschke:2017ejr} or a proxy thereof --- possibly preceded by a dynamical ``prehydrodynamic'' evolution~\cite{Schlichting:2019abc} and a subsequent transport ``afterburner'' ---, which describes well a large amount of the experimentally measured anisotropic-flow signals from collisions of heavy nuclei~\cite{Heinz:2013th}. 
However, the application of fluid dynamics to smaller systems or in peripheral collisions is more disputed~\cite{Weller:2017tsr,Zhao:2020pty}.
In addition, several approaches still compete for the early dynamics. 

A possible description of the system is given by kinetic transport theory, as we do in this paper. 
Within that model, it is possible to model (part of) the fireball evolution from the few-collisions regime to the hydrodynamic limit.
In this paper we investigate how the early-time evolution of several observables varies between these two extremes.

In the next section we introduce the tools we employ in this paper, which consist of a numerical transport code on the one hand and an analytical approach to the kinetic Boltzmann equation via a Taylor-series ansatz on the other hand.
Within both approaches we investigate several characteristics of a system of massless particles, in particular its anisotropic flow and spatial eccentricities, focusing on their development at early time (Sect.~\ref{sec:results}).
Finally we summarize our main findings and discuss them in Sect.~\ref{sec:discussion}.

\section{Methods and setup}
\label{s:methods}

In this section we briefly present the two methods used in our study, namely a numerical transport model (Sect.~\ref{subsec:numerical_simulations}) and an analytical approach to the kinetic Boltzmann equation (Sect.~\ref{subsec:analytical_approach}). 
Both require as starting point an initial distribution function, which we specify in Sect.~\ref{subsec:distribution_function}.

\subsection{Transport simulations}
\label{subsec:numerical_simulations}

To simulate the expansion of a system of relativistic degrees of freedom, we use on the one hand a numerical code implementing the covariant transport algorithm introduced in Ref.~\cite{Gombeaud:2007ub}.
As in Ref.~\cite{Roch:2020zdl}, to which we refer for further details, we consider massless (test) particles, modeled as hard spheres. 
Since we focus on characteristics of the system, namely its anisotropic flow and the evolution of the spatial eccentricities, which are mostly driven by transverse dynamics~---at least at the qualitative level~---, the system is purely two dimensional.
This restriction to the transverse plane allows us to overcome the statistical noise on the observables, in particular in the few collisions limit.

An advantage of the transport algorithm over the analytical calculations of Sect.~\ref{subsec:analytical_approach} is that one can smoothly cover the whole range from free-streaming to the hydrodynamic limit with a single model, by changing the cross section $\sigma$.  
To ensure covariance and locality, we make sure that the system remains dilute even in the fluid-dynamical regime: 
the parameter $D\equiv n^{-1/2}/\ell_\mathrm{mfp}$, which compares the relative sizes of the typical distance $n^{-1/2}$ between particles and the mean free path given by $\ell_\mathrm{mfp}\equiv 1/(n\sigma)$ --- both estimated in the initial state at the center of the system, where the particle density $n$ is maximum ---, is smaller than 0.1 for all our simulations.  

The computation time of the simulations typically grows as $N_{\mathrm{p}}^{3/2}$, where $N_\mathrm{p}$ denotes the number of test particles in the system.
To reduce statistical fluctuations, for every setup (initial geometry, Knudsen number) that we consider we perform $N_\mathrm{iter.}$ iterations of the simulation, which is equivalent to performing a single run with $N_\mathrm{p}\cdot N_\mathrm{iter.}$ test particles but computationally cheaper.
We typically use $N_{\mathrm{p}} = 2\times 10^5$ particles in the initial state and $N_\mathrm{iter.}={\cal O}(10^3)$, so that their product is always larger than $10^8$.

To allow a more accurate comparison with the analytical calculations presented in Sect.~\ref{subsec:analytical_approach}, which only account for the loss term of the kinetic Boltzmann equation, we also performed simulations with a ``$2\to 0$'' collision kernel. 
That is, two (test) particles that collide disappear from the system, which obviously violates every conservation law: energy, momentum, particle number. 
For those simulations we introduced labels ``active'' and ``passive'' for the test particles.
A collision can then only happen between two ``active'' particles.
After the collision, these are then labeled as ``passive'' for the remainder of the evolution and from that moment on they are no longer taken into account in the computation of any observable.

\subsection{Analytical calculations}
\label{subsec:analytical_approach}

A dilute system of particles undergoing binary collisions can also be described by a single-particle phase space distribution $f(t,\vec{x},\vec{p})$ obeying the kinetic Boltzmann equation with the appropriate $2\to 2$ collision term. 
In the regime of large Knudsen numbers, i.e.\ when the particles undergo very few rescatterings, one can expect that $f(t,\vec{x},\vec{p})$ will not depart much from a free-streaming distribution. 
This observation underlies a number of (semi-)analytical studies, in particular of anisotropic flow, in the few-collisions limit, using various approximations for the collision kernel~\cite{Heiselberg:1998es,Borghini:2010hy,Romatschke:2018wgi,Kurkela:2018ygx,Borghini:2018xum,Kurkela:2019kip,Kurkela:2020wwb,Kurkela:2021ctp,Ambrus:2021fej}.

Irrespective of any approximation, it was pointed out in Ref.~\cite{Borrell:2021cmh} that one can start from a Taylor expansion of the phase-space distribution at early times
\begin{align}
f(t,\vec{x},\vec{p}) = f^{(0)}(\vec{x},\vec{p}) &+ 
  t\, \partial_t f(t,\vec{x},\vec{p})\big|_0 \cr 
  & + \frac{t^2}{2} \partial_t^2 f(t,\vec{x},\vec{p})\big|_0  + \cdots,
\label{eq:Taylordg}
\end{align}
where $f^{(0)}$ denotes the initial distribution while the successive time derivatives are evaluated at the initial time $t=0$. 
By making use of the relativistic Boltzmann equation, rewritten in the form
\begin{equation}
 \partial_t f(t,\vec{x},\vec{p}) = 
 -\frac{\vec{p}}{E}\cdot\vec{\nabla}_{\!x} f(t,\vec{x},\vec{p}) + {\cal C}[f],
 \label{eq:be}
\end{equation}
one can replace every time derivative in Eq.~\eqref{eq:Taylordg}, so that given the form of the collision kernel, the whole evolution of $f(t,\vec{x},\vec{p})$ is governed by the initial distribution and its spatial derivatives~\cite{Borrell:2021cmh}:
\begin{widetext}
\begin{align}
f(t,\vec{x},\vec{p}) = f_{\mathrm{f.s.}}(t,\vec{x},\vec{p}) &+ 
  t\,{\cal C}[f]\big|_0  + 
  \frac{t^2}{2} \bigg(\!\!-\!\frac{\vec{p}}{E}\cdot\vec{\nabla}_{\!x} {\cal C}[f] +
   \partial_t {\cal C}[f]\bigg)_{\!0} \cr
   &+\frac{t^3}{3!}\bigg( \frac{\big(\vec{p} \cdot \vec{\nabla}_{\!x}\big)^2}{E^2} {\cal C}[f] - 
 \frac{\vec{p}}{E}\cdot\vec{\nabla}_{\!x} \partial_t {\cal C}[f] +
 \partial_t^2 {\cal C}[f] \bigg)_{\!0} + \mathcal{O} (t^4),
 \label{FullExpansion}
\end{align} 
\end{widetext}
where $f_{\mathrm{f.s.}}$ is obtained by grouping all terms which do not contain the collision kernel and is therefore the free-streaming distribution with the same initial condition at $t=0$.
Again, the time derivatives of the collision kernel can be re-expressed using the Boltzmann equation, so that only $f^{(0)}$ and its spatial derivatives are involved. 
Note that these time derivatives of ${\cal C}[f]$ actually involve increasing powers of the cross section characterizing the rescatterings~\cite{Borrell:2021cmh}, as will be illustrated in the following section.

From this point, one can in principle compute any quantity at early times. 
Obviously if more orders in time are included, the result will be closer to the full solution, although it will always depart at later times when only a finite number of powers of $t$ are taken into account.
To match the numerical simulations, we shall make all analytical calculations in 2 dimensions and with massless particles.

A drawback of the analytical approach is that the gain term of the usual $2\to 2$ collision kernel of the Boltzmann equation is difficult to handle.
In contrast to the loss term, the ``particle of interest'' --- that with the same momentum as appears on the left hand side of the Boltzmann equation --- does not enter it directly, and the form of the differential cross section does matter. 
To bypass this issue, here we only consider the loss term of the collision integral, i.e.\ a $2 \to 0 $ collision kernel~\cite{Heiselberg:1998es,Borghini:2010hy}
\begin{align}
\label{C[f]_2->0}
{\cal C}_{2\to 0}[f] = -\frac{E}{2}\int\! f({\bf p})f({\bf p}_1)_{} v_{\mathrm{rel.}}
  \sigma\,{\rm d}^2\,{\bf p}_1,
\end{align}
with $v_{\rm rel.}$ the M{\o}ller velocity between the two colliding particles, while for brevity we did not denote the time and position variables.
This collision term clearly does not conserve energy, momentum or particle number, but on the other hand we are able to push our calculations to higher order in the total cross section $\sigma$. 

As stated above, we also implement a $2\to 0$ collision kernel in the numerical code by ignoring particles that have already undergone a collision.
A remaining important difference between the analytical calculations and the simulations is the order in $\sigma$.
While the analytical calculations contain only a finite amount of orders due to the Taylor expansion, the numerical simulations contain the total cross section to all orders.

\subsection{Initial distribution function}
\label{subsec:distribution_function}

In this paper, similar in that respect to other studies in the literature~\cite{Heiselberg:1998es,Borghini:2010hy,Romatschke:2018wgi,Kurkela:2018ygx,Kurkela:2019kip,Kurkela:2020wwb,Kurkela:2021ctp,Ambrus:2021fej}, we use a simplified semi-realistic geometry for the initial state for both our numerical and analytical investigations. 

We assume that the initial phase space distribution $f^{(0)}$ factorizes like
\begin{align}
f^{(0)}({\bf x},{\bf p}_{\mathrm{T}}) = F({\bf x})\, G({\bf p}_\mathrm{T},T({\bf x})),
\label{InitialDistribution}
\end{align}
where the spatial part $F$ governs the geometry, while the momentum part $G$ takes the form of a thermal distribution with a temperature that can depend on position:
\begin{align}
G({\bf p}_\mathrm{T}) = \frac{1}{2\pi T({\bf x})^2}e^{-|{\bf p}_\mathrm{T}|/T({\bf x})}.
	\label{eq:momentum_dist}
\end{align}
Note that $G$ is normalized to unity, so that $F({\bf x})$ is actually the particle-number density.
The temperature is determined by the latter via
\begin{align}
T({\bf x}) \propto \sqrt{F({\bf x})},
\label{eq:T_match}
\end{align}
following the equation of state of a perfect gas of massless particles in two dimensions.
This relation implies that the central region will have a higher temperature than the outer ones. 

Using polar coordinates in the transverse plane, the profile in position space is taken to be 
\begin{align}
F(r,\theta) = \frac{N_{\rm p}\,{\rm e}^{-\frac{r^2}{2R^2}}}{2\pi R^2}
\left[1-\sum_{k=2}^3 \tilde{\varepsilon}_k {\rm e}^{-\frac{r^2}{2R^2}}\left(\frac{r}{R}\right)^{\!\!k}\cos(k\theta)\right],
\label{eq:initial_distribution}
\end{align}
where $N_{\rm p}$ is the total initial number of particles and $R$ a characteristic system size, which sets the typical time scale of the system evolution. 
The value of $R$ (6.68\;fm in the calculations) alone is irrelevant for the results, it is only meaningful when combined with $\sigma$ (which in two dimensions has the dimension of a length) and $N_{\rm p}$ to yield a dimensionless quantity like the Knudsen number.
Consistent with the normalization of the momentum distribution $G$, $F$ is normalized to $N_{\rm p}$, so that the initial phase-space distribution is also normalized to the total number of particles, as it should. 

The two (real) parameters $\tilde{\varepsilon}_2$ and $\tilde{\varepsilon}_3$ give the degree of asymmetry of the initial geometry: a straightforward integral gives for the eccentricities~\eqref{eccentricity} $\varepsilon_2^{\bf x} = \tilde{\varepsilon}_2/4$ and $\varepsilon_3^{\bf x} = \tilde{\varepsilon}_3/\sqrt{2\pi}$ with $\Phi_2 = \Phi_3 = 0$, where the latter choice was made for the sake of simplicity, without any impact on our results.\footnote{If $\tilde{\varepsilon}_n$ is taken to be complex, then the symmetry-plane angle $\Phi_n$ is the argument of $\tilde{\varepsilon}_n$.}
All results shown in the following were obtained with initial profiles such that only one eccentricity $\tilde{\varepsilon}_2$ or $\tilde{\varepsilon}_3$ is non-zero --- up to numerical fluctuations. 
Our standard choice for the non-vanishing $\tilde{\varepsilon}_n$ is that it yields $\varepsilon_n^{\bf x}\simeq 0.15$.

\begin{figure*}[!ht]
	\includegraphics[width=0.495\linewidth]{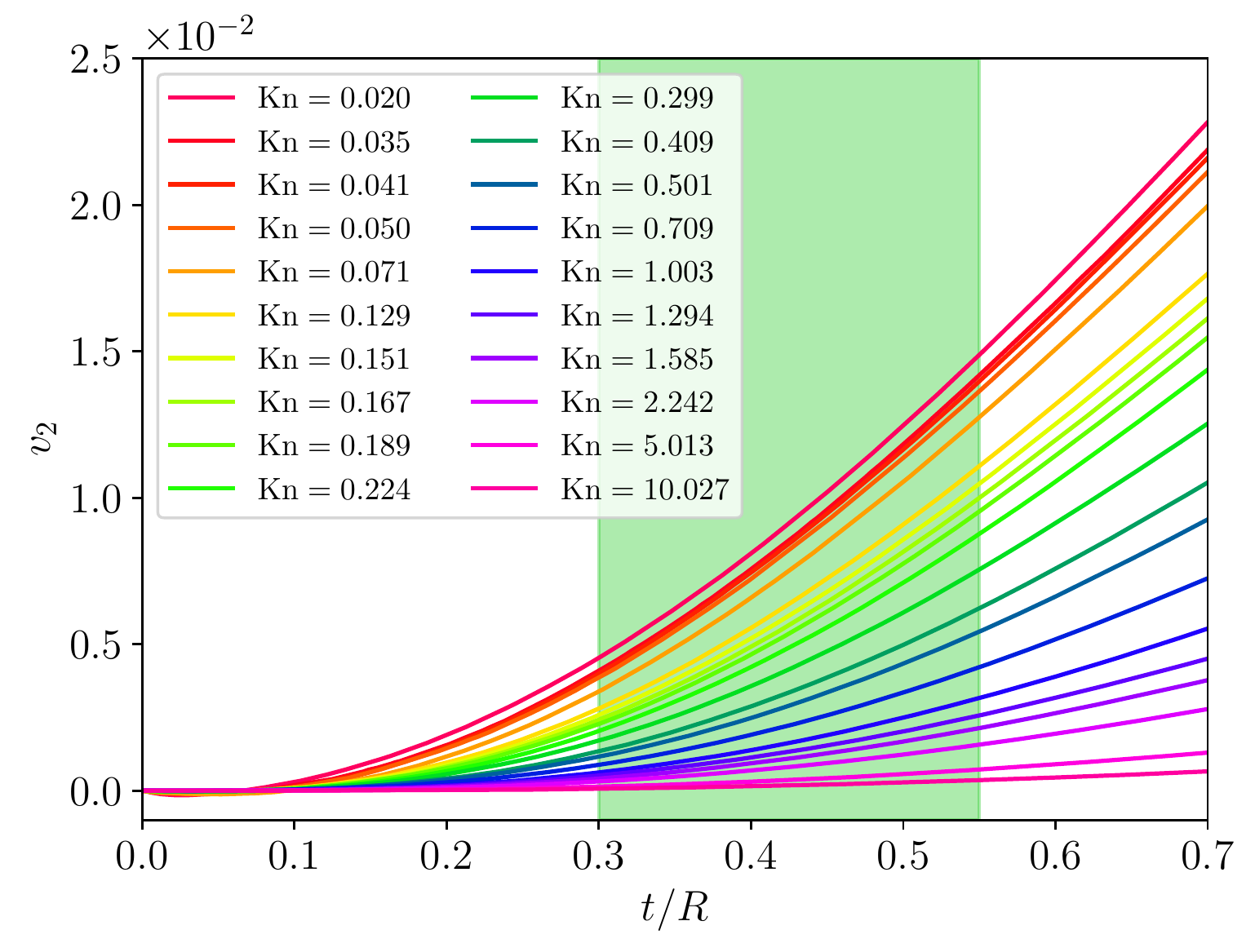}
	\includegraphics[width=0.495\linewidth]{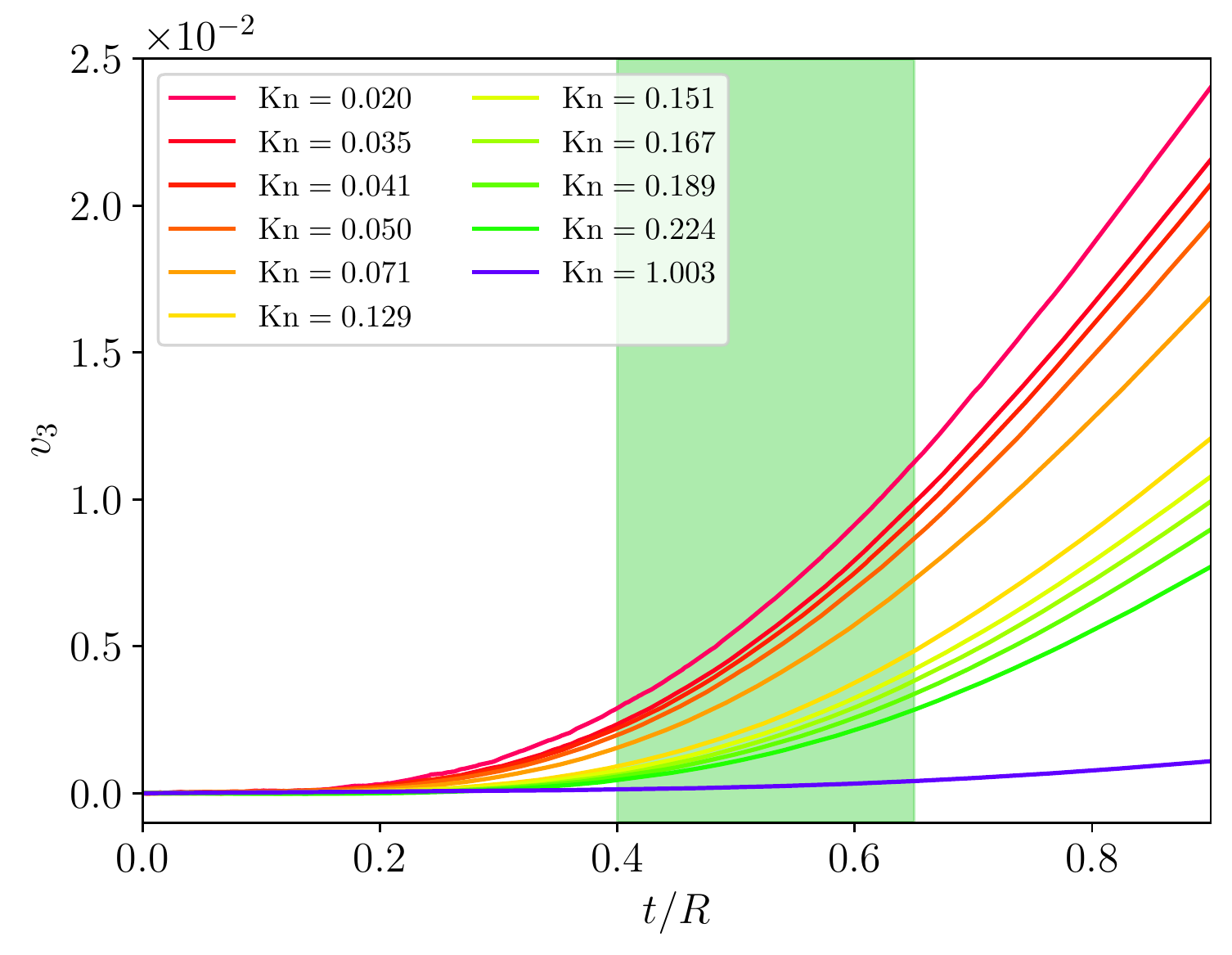}\vspace{-3mm}
	\caption{$v_2$ (left) and $v_3$ (right) as function of $t/R$ for different Knudsen numbers. The green shaded regions mark the interval for the upper end points of the fits with Eq.~\eqref{vn_fit-function}.}
	\label{fig:loglog_flow}
\end{figure*}

An important feature is that the initial distribution~\eqref{InitialDistribution} is isotropic in momentum space. 
This is a necessary ingredient for the analytical calculations and it is mostly fulfilled by the transport simulations, up to numerical fluctuations that we now discuss.
In the initial state of the transport simulation, $N_\mathrm{p}$ test particles are first sampled from the particle number density~\eqref{eq:initial_distribution}, and their momenta are sampled from the Boltzmann distribution~\eqref{eq:momentum_dist} with the appropriate local temperature.
Due to the finite number of particles, the resulting initial (transverse) momentum distribution, integrated over all positions, is not exactly isotropic, but shows small anisotropic flow coefficients of order $1/\sqrt{2N_\mathrm{p}}$~\cite{Voloshin:1994mz}.
Discarding the initial $v_1$ is easy, by subtracting $1/N_\mathrm{p}$ times the total momentum of the system from the momentum of every particle.\footnote{Strictly speaking, this slightly deforms the momentum distribution~\eqref{eq:momentum_dist}, but only minimally, as shown by the agreement with analytical calculations using Eq.~\eqref{eq:momentum_dist}.}
However, setting the other initial Fourier coefficients to zero is not so easy. 
To circumvent the problem, we perform $N_\mathrm{iter.}$ iterations with exactly the same initial geometry, i.e.\ the positions of the $N_\mathrm{p}$ test particles are unchanged, but with different samplings of the momentum distribution. 
The results of the $N_\mathrm{iter.}$ simulations with the same geometry are then averaged, reducing fluctuations in particular in the initial state by a factor $\sqrt{N_\mathrm{iter.}}$. 
As stated in Sect.~\ref{subsec:numerical_simulations}, this is (at least to a good approximation) equivalent to performing a single simulation with $N_\mathrm{p}\cdot N_\mathrm{iter.}$ test particles, but computationally significantly cheaper.
In our simulations, the initial $v_n$ values after averaging over iterations is close to zero within the expected numerical uncertainty. 
Accordingly we shall systematically shift our anisotropic-flow curves to start at 0, adding at $t=0$ an error bar indicating the numerical fluctuation $1/\sqrt{2N_{\rm p}N_{\rm iter.}}$.

We characterize the amount of rescatterings in the evolution via the Knudsen number $\mathrm{Kn}\equiv\ell_\mathrm{mfp}/R$, where the particle mean free path $\ell_\mathrm{mfp}$ is estimated at the center of the system --- where the particle density is largest --- in the initial state.
A small Knudsen number means a large mean number of rescatterings per particle, corresponding to the ``fluid dynamical limit'', while a large Kn means a system with few rescatterings: by construction, Kn is inversely proportional to the cross section $\sigma$.

\section{Results}
\label{sec:results}

In this section we present our results, starting in Sect.~\ref{subsec:flow_coefficients_Kn} with the change in the early-time scaling behavior of anisotropic flow as one goes from the few collisions regime to the hydrodynamic limit. 
In the following subsections we focus on the few collisions regime and compare the numerical and analytical approaches for the early time behaviors of the number of rescatterings (Sect.~\ref{subsec:number_particles}), anisotropic flow (Sect.~\ref{subsec:flow_coefficients}), and eventually spatial eccentricities and other geometric quantities (Sec.~\ref{subsec:eccentricities}). 
Eventually, we discuss alternative observables to quantify the momentum anisotropies in Sect.~\ref{subsec:alternative_flow_observables}, again across the whole range of Knudsen numbers and comparing with analytical calculations when Kn is large.

\subsection{Onset of anisotropic flow from small to large Knudsen number}
\label{subsec:flow_coefficients_Kn}

Let us first investigate how the early-time development of anisotropic flow --- here quantified via the Fourier harmonics~\eqref{vn_def} of the particle transverse-momentum distribution, as is most often done --- evolves across Knudsen numbers for a fixed initial geometry. 
As is customary in transverse flow studies~\cite{Sorge:1996pc,Heinz:2001xi}, ``early times'' are to be understood in comparison to the typical transverse size of the system $R$, say roughly $t/R$ of order 0.1--0.5.

As is well known and will be shown again below, the value of $v_n$ at a fixed final time increases with the number of rescatterings in the system, when going from the free-streaming limit ${\rm Kn}\to\infty$ --- in which no anisotropic flow develops --- to the fluid-dynamical limit ${\rm Kn}\to 0$. 
Throughout this paper we are not so much interested in that known behavior, but rather in the onset of $v_n(t)$, namely its dependence on $t$ at early times. 
Let us quickly recall the findings in the literature: 
In fluid dynamics, numerical simulations~\cite{Heinz:2002rs,Kolb:2002cq,Kolb:2003dz,Teaney:2010vd} or general scaling arguments~\cite{Vredevoogd:2008id} yield $v_2(t) \propto t^2$ and more generally $v_n(t) \propto t^n$ at early times. 
In contrast, transport simulations~\cite{Gombeaud:2007ub,Alver:2010dn} or analytical studies~\cite{Heiselberg:1998es,Borghini:2010hy} in the few-rescatterings regime yield the slower growth $v_n(t) \propto t^{n+1}$. 
Here we wish to bridge the gap between the two regimes, looking at both elliptic flow $v_2$ and triangular flow $v_3$. 
In this subsection, we use the normal $2\to 2$ collision kernel in our simulations. 

\begin{figure*}[!t]
	\includegraphics[width=0.495\linewidth]{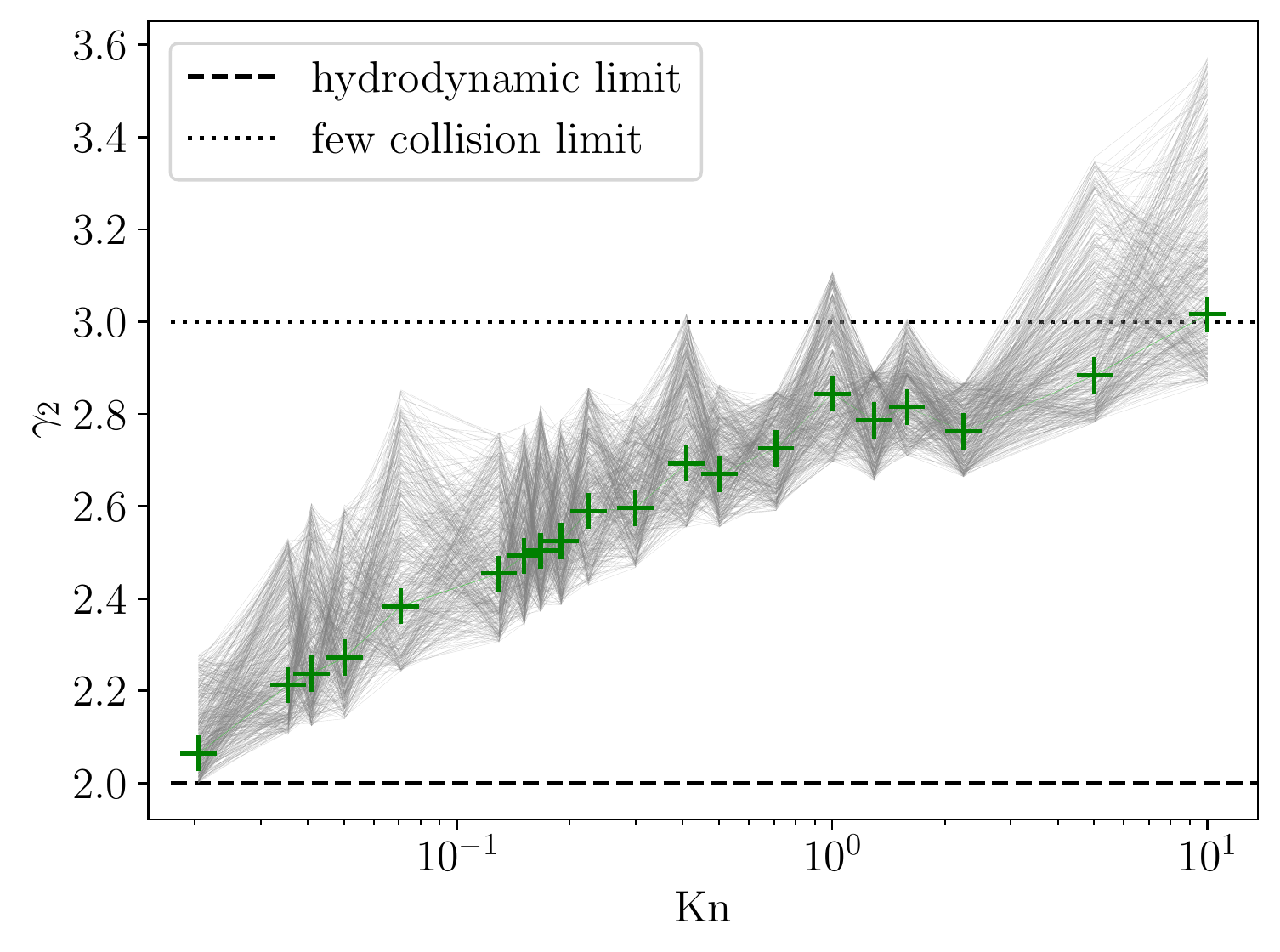}
	\includegraphics[width=0.495\linewidth]{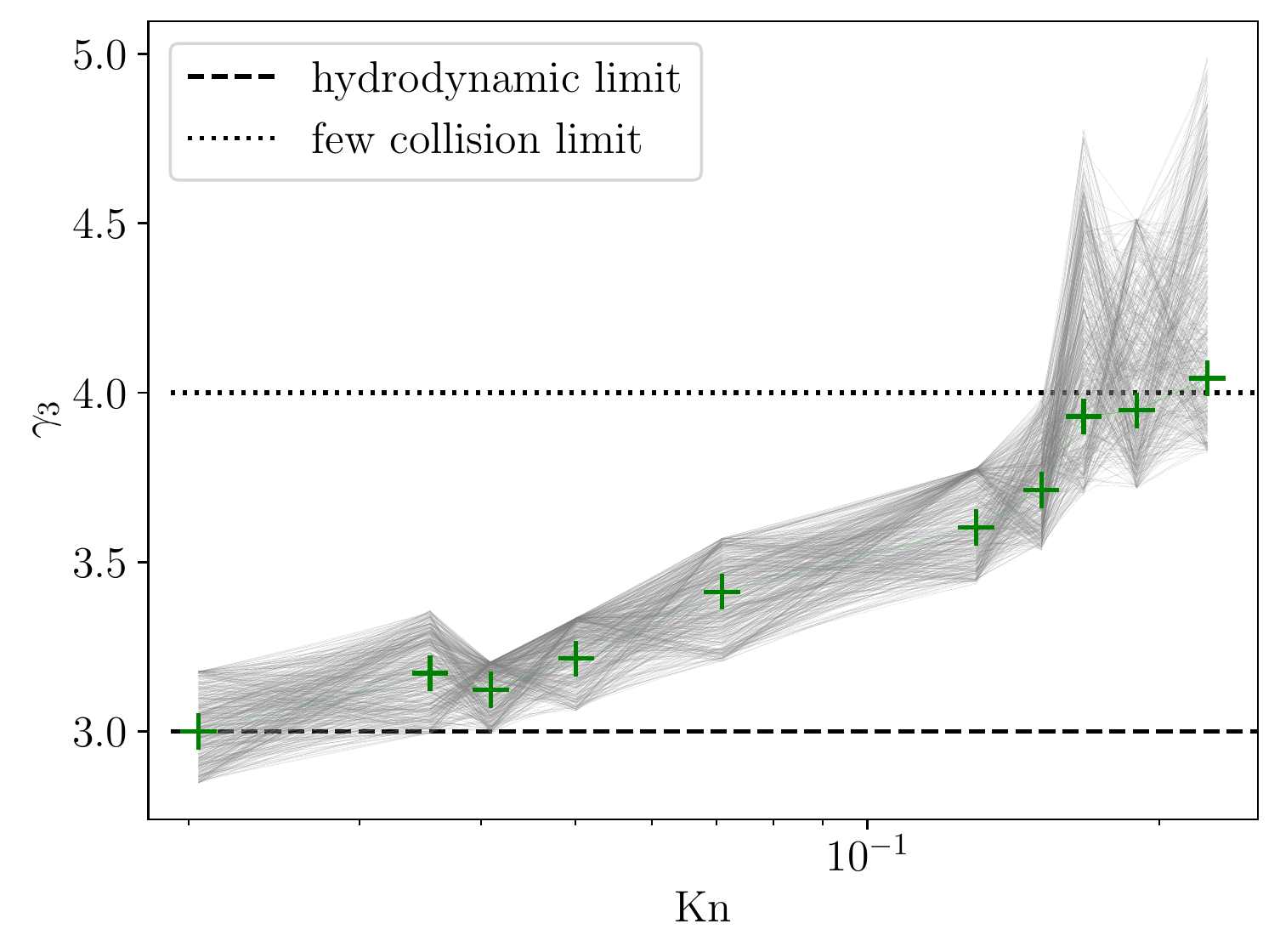}\vspace{-3mm}
	\caption{Scaling exponent $\gamma_2$ (left) and $\gamma_3$ (right) as a function of the Knudsen number. 
		The gray lines indicate the 500 realizations of the fit over different time intervals, while the green points correspond to the weighted average~\eqref{eq:weighted_av}.}
	\label{fig:gamma}
\end{figure*}

In Fig.~\ref{fig:loglog_flow} we display $v_2(t)$ and $v_3(t)$ at early times, namely for $t\leq 0.7R$ resp.\ $t\leq 0.9R$, where $R$ is the typical system size, for various Knudsen numbers ranging from ${\rm Kn} = 0.02$ to ${\rm Kn} \approx 10$. 
The latter amounts to less than 0.1 rescatterings per particle on average, while in the former case the particles rescatter about 25 times on average, corresponding to the fluid-dynamical regime. 
As expected, for larger Knudsen numbers the spatial asymmetry is less efficiently translated into a momentum anisotropy.
Note also that for a fixed $t$ and a given Kn the value of $v_3$ is significantly smaller than that of $v_2$, which is why we restrict ourselves to a smaller range in Kn.

Inspired by the known results, we fit the onset of those curves with a power-law ansatz 
\begin{equation}
v_n\bigg(\frac{t}{R}\bigg) =
  \beta_n\bigg(\frac{t}{R}\bigg)^{\!\!\gamma_n}.
  \label{vn_fit-function}
\end{equation}
To account for the small initial flow value due to the finite test particle number, we shifted all curves to zero at $t=0$.
Since we are only interested in the scaling exponent $\gamma_n$ this shift does not influence our result, yet has the benefit that we do not have to introduce an extra offset parameter in the fit routine.
More physical information is contained in the parameters $\beta_n$ and $\gamma_n$ which are functions of the Knudsen number.
In fact, it is somewhat intuitive that $\beta_n$ should increase with decreasing Kn, which is what we indeed found.

A somewhat problematic issue is the size of the interval in $t/R$ over which we should perform the fit. 
The lower end point of the fitting interval is always taken at $t=0$. 
Regarding the upper end point of the interval, on the one hand it should not be too large to make sure that we are still fitting early-time behavior. 
On the other hand, this final end point should not be too small either, otherwise the signal to be fitted is too small and may still be contaminated by numerical noise. 
Accordingly, we defined a range of values, represented by the green bands in Fig.~\ref{fig:loglog_flow}, for the maximum time of the fitting interval: $0.3\leq t_{\max}/R\leq 0.55$ for $v_2$ resp.\ $0.4\leq t_{\max}/R\leq 0.65$ for $v_3$. 
We randomly chose 500 values for $t_{\max}$,\footnote{We drew $t_{\max}$ from a uniform distribution over the respective range of values.} and used them to perform fits with the power law~\eqref{vn_fit-function} for all values of Kn.
In Fig.~\ref{fig:gamma} we show for $n=2$ and $n=3$ the exponents $\gamma_n$ of these fits, joining by gray lines the values obtained at different Knudsen numbers but with the same value of $t_{\max}$.
We then performed a weighted average of the exponents $\gamma_n$ over the 500 different realizations of the fit interval, giving less weight to the fits with larger uncertainties by using
\begin{equation}
\bar{\gamma}_n = \frac{\sum_j \gamma_{n,j}/\sigma^2_{\gamma_n,j}}{\sum_j 1/\sigma^2_{\gamma_n,j}},
\label{eq:weighted_av}
\end{equation}
where $\sigma^2_{\gamma_n,j}$ is the (squared) uncertainty on $\gamma_{n,j}$ given by the fitting routine for the $j$-th realization, while $j$ runs over the 500 realizations.
Performing the 500 realizations of the fit gives a better insight on the range of possible exponents at each Knudsen number and provides a very conservative error band for the uncertainty on the exponent, which is actually not symmetric about the average value~\eqref{eq:weighted_av}.

Even accounting for the uncertainty bands, $\gamma_2$ (top panel of Fig.~\ref{fig:gamma}) and $\gamma_3$ (bottom panel) show a clear trend, namely a crossover from the hydrodynamic ($\gamma_n = n$) to the few-collisions regime ($\gamma_n = n+1$) over some range in the Knudsen number.
Comparing the horizontal axes of the two plots, we see that the transition takes place over a much smaller Kn-range for $v_3$. 
Moreover, the change in the early-time evolution of $v_3$ takes place for a typical Kn for which $v_2$ is already close to its fluid-dynamical behavior. 

\begin{figure}[!t]
	\centering
	\includegraphics[width=\columnwidth]{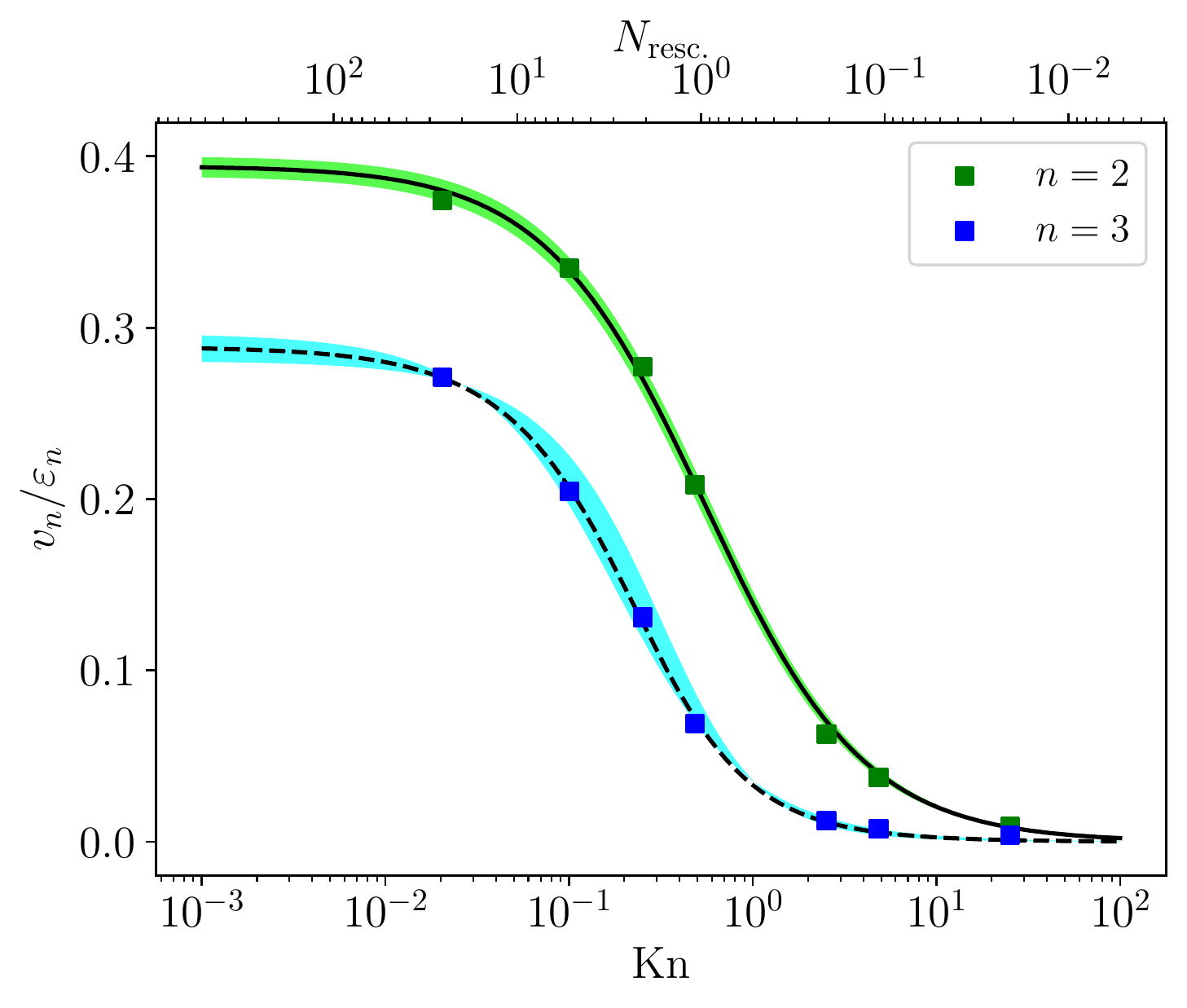}
	\caption{$v_n/\varepsilon_n$ as a function of the Knudsen number and the mean number of rescatterings $N_\mathrm{resc.}$ per particle, where $v_n$ and $N_\mathrm{resc.}$ are estimated at time $t=100R$. The fits were performed with Eqs.~\eqref{eq:v2_Kn}--\eqref{eq:v3_Kn}. The colored bands show the $1\sigma$ fit error.}
	\label{fig:vn_vs_Kn}
\end{figure}
The latter finding parallels the already known different behaviors of the ``final'' values of $v_2$ and $v_3$ --- that is, once $v_n(t)$ has reached its maximum value and evolves only little --- with varying Knudsen number:
starting from the free-streaming limit and decreasing Kn, the $v_2$ signal rises before that of $v_3$~\cite{Alver:2010dn,Roch:2020zdl}, as we also show in Fig.~\ref{fig:vn_vs_Kn} for the specific initial conditions used in this paper.
The fits to the points were performed with~\cite{Bhalerao:2005mm}
\begin{equation}
v_2=\frac{v_2^{\mathrm{hydro}}}{1+\mathrm{Kn}/\mathrm{Kn}_0} 
    \label{eq:v2_Kn}
\end{equation}
and~\cite{Alver:2010dn}
\begin{equation}
v_3=\frac{v_3^{\mathrm{hydro}}(1+B_3\mathrm{Kn})}{1+(A_3+B_3)\mathrm{Kn}+C_3\mathrm{Kn}^2},
    \label{eq:v3_Kn}
\end{equation}
where the parameter $v_n^\mathrm{hydro}$ is the value in the ideal fluid-dynamical limit.
That $v_3$ departs from its few-rescatterings behavior at a smaller Kn than $v_2$ can readily be explained at least on the qualitative level: 
At fixed system size $R$, a triangular structure, as will give rise to $v_3$, involves a smaller typical ``wavelength'' (in the azimuthal direction) than an elliptic shape. 
To resolve this smaller wavelength, a smaller mean free path is needed, i.e.\ a smaller Knudsen number.\footnote{An alternative argument in the language of fluid dynamics was given in Ref.~\cite{Alver:2010dn}.}

As a final remark, note that fitting the slow early-time growth of triangular flow in the few-collisions regime, $v_3(t)\propto t^{\gamma_3}$ with $\gamma_3\approx 4$, appears to be at the limit of what we can do with reasonable uncertainties. 
This is why we did not try to investigate the early time behavior of $v_4$, which should grow as $t^4$ in the hydrodynamic limit and slower when going to larger Knudsen numbers.

\subsection{Number of rescatterings}
\label{subsec:number_particles}

In the few-rescatterings regime we have expansion~\eqref{FullExpansion} at our disposal, with which one can compute the early-time behavior of observables. 
Keeping only the first non-vanishing contribution from the Taylor expansion to a given observable, in particular $v_n(t)$, one can hardly expect to obtain a non-integer scaling exponent like the $\gamma_n$ found at ``intermediate'' Knudsen numbers in the previous subsection. 
However, we can investigate whether summing the contributions from several terms in the Taylor expansion could lead to an effective behavior similar to that of the full transport simulations.

We begin with the number of rescatterings in the system, scaled by twice the number of particles $N_{\rm p}$ in the initial state, where we include the factor 2 since each collision involves two particles.
In analytical calculations, this number of rescatterings per particle $N_{\rm resc.}$ is simply one half of the integral over time  of the collision rate, i.e.\ it can be deduced from the integral of the $2\to 0$ collision kernel~\eqref{C[f]_2->0} over phase space:
\begin{align}
N_{\rm resc.}(t) &= \frac{1}{2N_{\rm p}}\!\int_0^t \Gamma(t')\,{\rm d} t' \cr
 &= \frac{1}{N_{\rm p}}\!\int_0^t \!\int\! -\mathcal{C}_{2\to 0}[f(t',{\bf x},{\bf p})]\,
 \frac{{\rm d}^2{\bf x}\,{\rm d}^2{\bf p}}{E}\,{\rm d} t'.\qquad
 \label{Nresc_analytical}
\end{align}
Note that this formula also holds for a system undergoing any kind of binary collisions between two identical particles, despite the apparently restrictive notation $2\to 0$. 

Using expansion~\eqref{FullExpansion} one can compute the phase-space distribution at time $t'$ entering Eq.~\eqref{Nresc_analytical}.
As stated in Sect.~\ref{subsec:analytical_approach}, we were able to calculate $f(t,{\bf x},{\bf p})$ at early times only in the $2\to 0$ scenario, which is why we now consider also that model --- in which the total particle number is not conserved --- in the numerical simulations. 
On the analytical side, we computed $N_{\rm resc.}$ up to order $t^{11}$, considering terms up to order $\sigma^5$ in the cross section.\footnote{Strictly speaking, at ${\cal O}(t^{11})$ one finds terms up to order $\sigma^{10}$, so our $N_{\rm resc.}(t)$ at order $t^{11}$ is incomplete. Every new order in $\sigma$ means a significant increase in the number of terms to be calculated, which is why we were slightly inconsistent here.}

\begin{figure}[!t]
	\centering
	\includegraphics[width=\columnwidth]{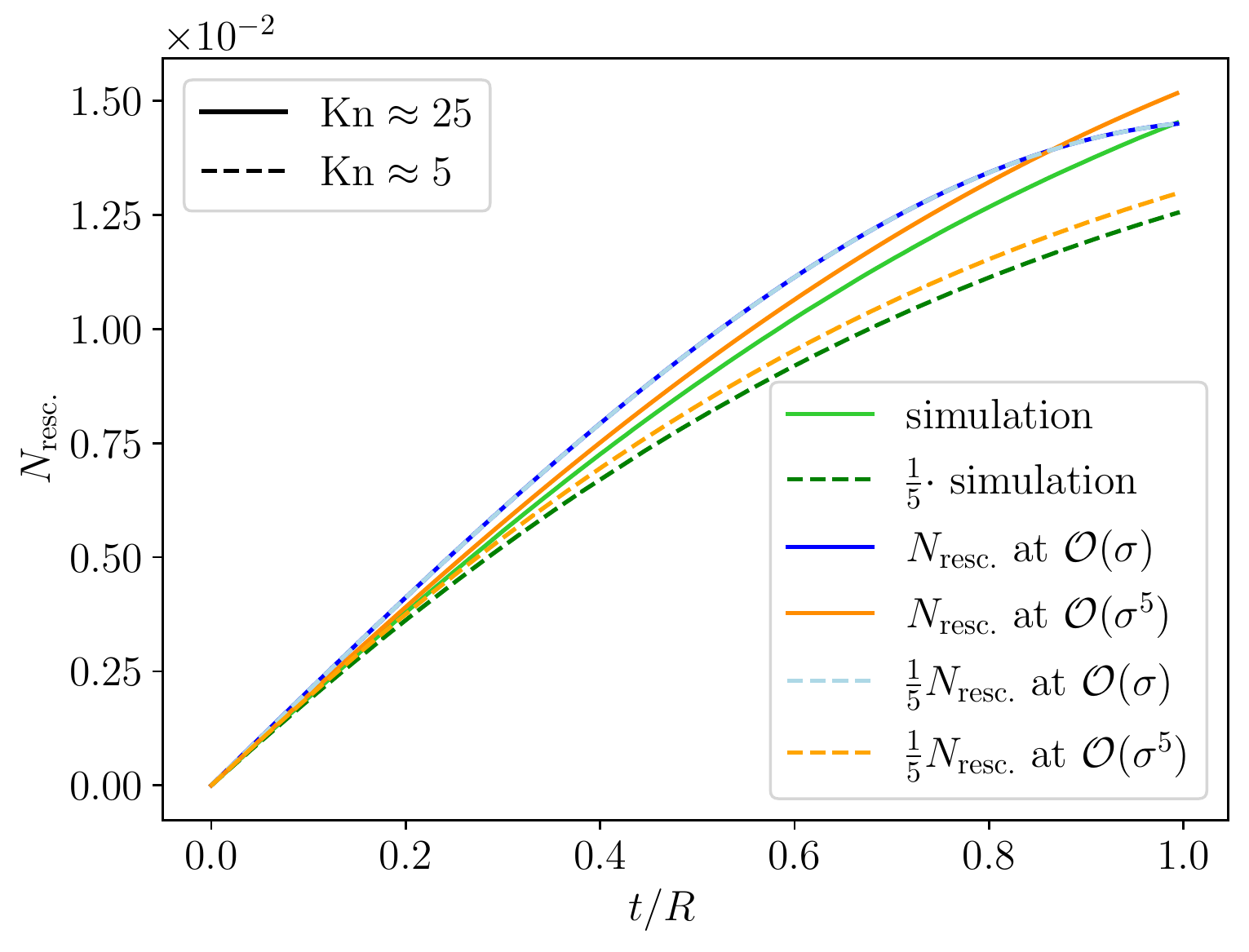}
	\caption{Number of rescatterings per particle for $\mathrm{Kn}\approx 25$ (full) and $\mathrm{Kn}\approx 5$ (dashed) for the $2\rightarrow 0$ collision kernel simulation (green). The analytical results --- computed up to ${\cal O}(t^{11})$ --- are shown at $\mathcal{O}(\sigma)$ (blue) and $\mathcal{O}(\sigma^5)$ (orange) for the corresponding Knudsen numbers.}
	\label{fig:Nresc}
\end{figure}

The importance of going beyond linear order in $\sigma$ is visible in Fig.~\ref{fig:Nresc}. 
There we show in green the results of simulations at a Knudsen number of order 25 (full line) at a ${\rm Kn}\approx 5$ (dashed line), where the latter are scaled by a factor $1/5$: if $N_{\rm resc.}(t)$ was simply proportional to $\sigma$, the two curves would coincide, which is clearly not the case. 

We also display in blue the computed $N_{\rm resc.}(t)$ at order $\sigma$: the curves at ${\rm Kn}\approx 25$ and ${\rm Kn}\approx 5$ do coincide, as they should, and they are reasonably close to the value of numerical simulations at ${\rm Kn}\approx 25$ up to $t/R \approx 0.3$. 
The agreement between analytical and numerical results is further increased when going up to order $\sigma^5$ (in orange) for both values of Kn we consider here. 

Overall the number of rescatterings can thus be well described in the analytical approach. 
This is a necessary ingredient in our study since other observables of more complicated nature, as e.g.\ $v_n(t)$, depend in a way or another on the number of rescatterings.
Note that ${\rm Kn}\approx 5$ means that about 16\% of the initially present particles disappear during the whole system evolution, most of them in the initial stage shown in Fig.~\ref{fig:Nresc}. 
Accordingly, the results of $2\to 2$ and $2\to 0$ simulations differ significantly over that time interval, which is why it is necessary to compare the latter with our (equally $2\to 0$) analytical approach.

\subsection{Anisotropic flow coefficients}
\label{subsec:flow_coefficients}

Let us come back to our starting point, namely the behavior of anisotropic flow harmonics $v_n$ at early times, in a system without ``preflow'', i.e.\ $v_n(t=0) = 0$.
As in the case of the number of rescatterings and in contrast to Sect.~\ref{subsec:flow_coefficients_Kn}, here we compute the flow coefficients in the $2\to 0$ scenario, as was done in Refs.~\cite{Heiselberg:1998es,Borghini:2010hy}.
It turns out that at large Kn numerical simulations with the full $2\to 2$ kernel or the $2\to 0$ version yield very similar results for $v_2(t)$ as long as $t\lesssim R$.\footnote{The signal in the $2\to 0$ setup is slightly smaller than in the $2\to 2$ case, and the deviation between both models increases with decreasing Kn, which is due to the faster dilution in the $2\to 0$ system.
	For a more in-depth study of the differences between the $2\to 2$ and $2\to 0$ collision kernels, including also the late time behavior of the flow coefficients, we refer to Ref.~\cite{Bachmann:2022cls}.}
Note that it is hardly possible to study the dependence of $v_n$ over the full Kn range as in Figs.~\ref{fig:gamma} und \ref{fig:vn_vs_Kn}, but only at large Kn, since ${\rm Kn} = 1$ means that already about 60--80\% of the initially present particles disappear. 

\begin{figure}[!t]
    \centering
    \includegraphics[width=\columnwidth]{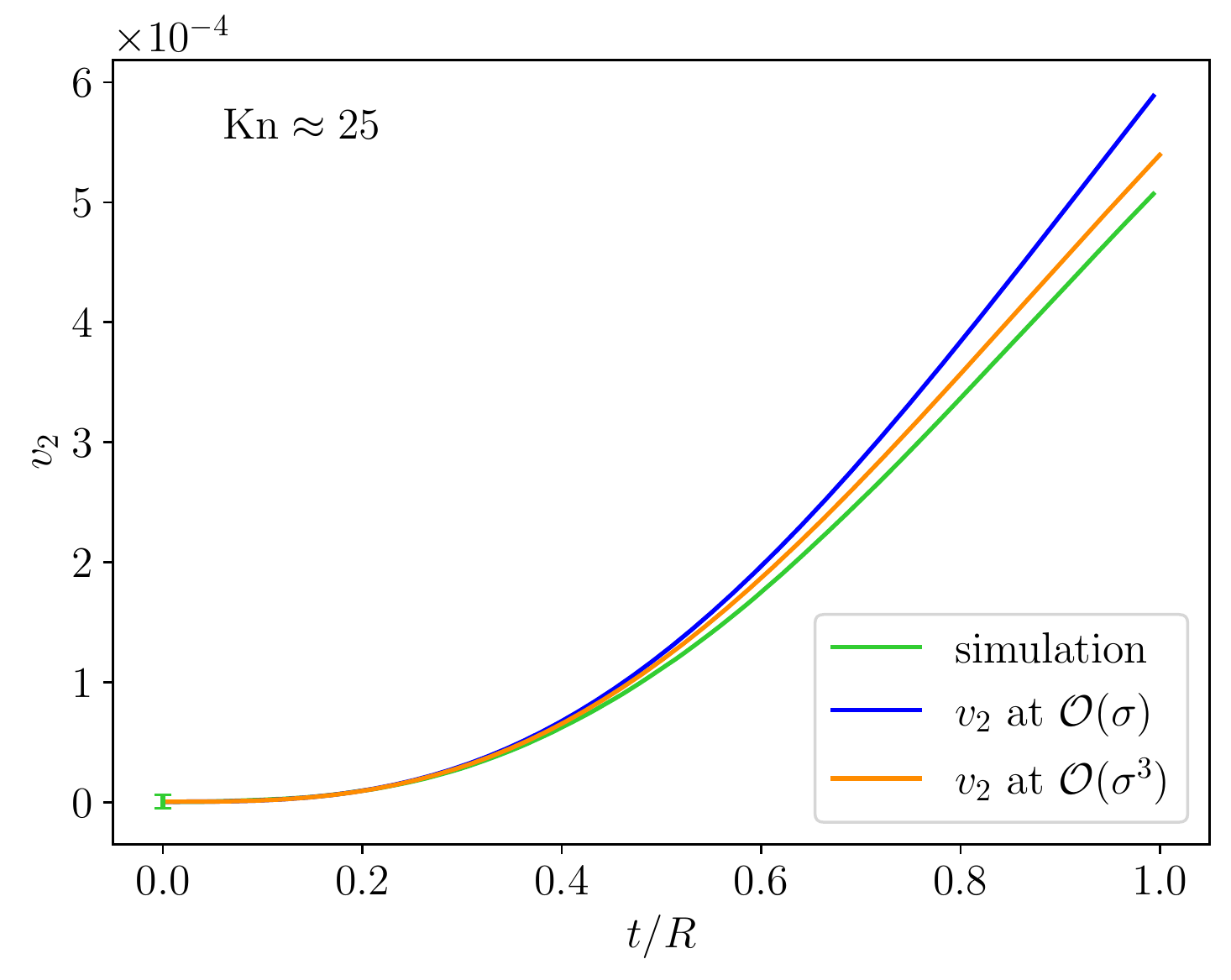}
    \includegraphics[width=\columnwidth]{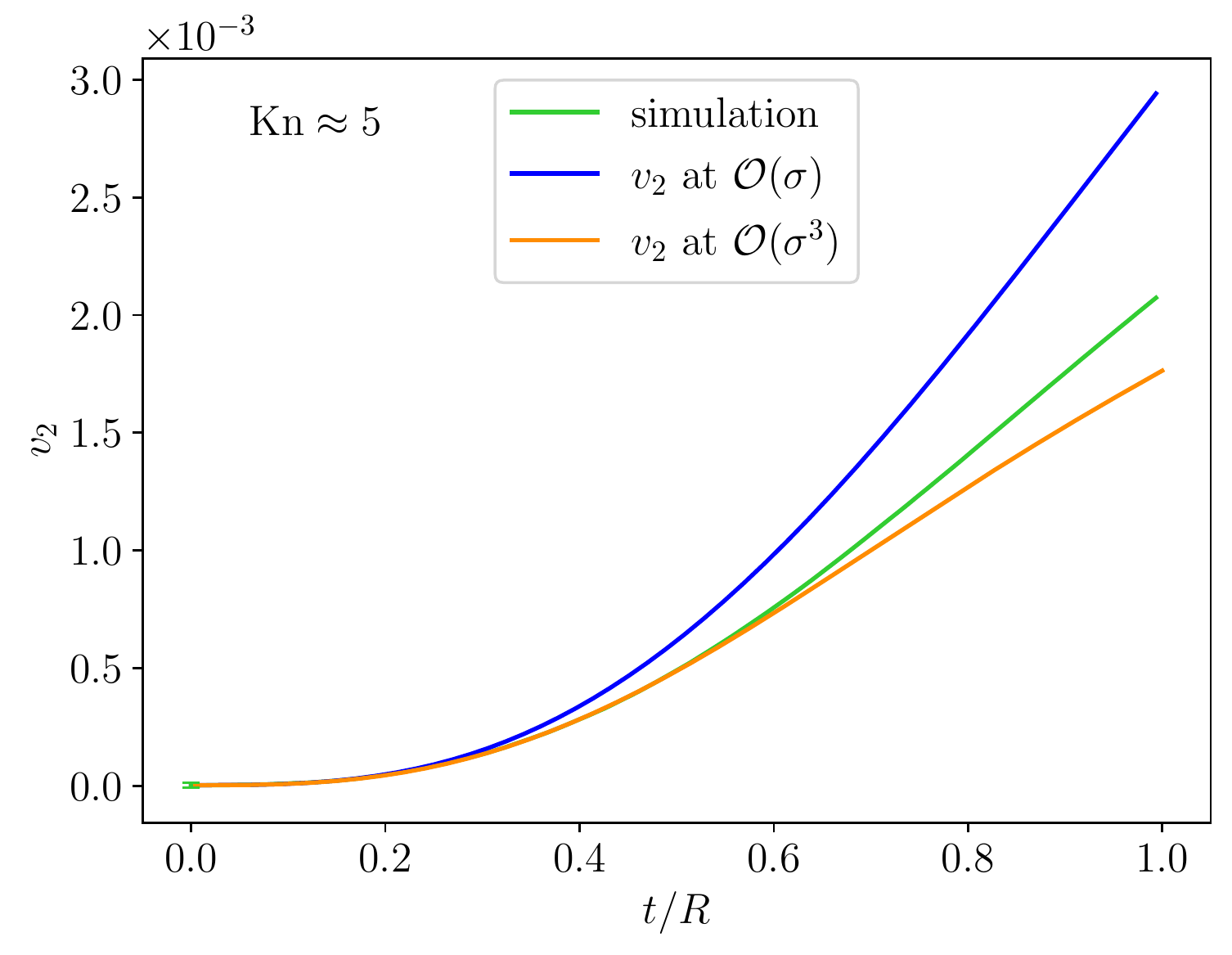}
    \caption{Elliptic flow $v_2$ as a function of $t/R$ for $\mathrm{Kn}\approx 25$ (top) and $\mathrm{Kn}\approx 5$ (bottom). Results from the numerical $2\to 0$ simulations are shown in green and the analytical results are shown to $\mathcal{O}(\sigma)$ (blue) and $\mathcal{O}(\sigma^3)$ (orange). Both analytical curves include terms up to $\mathcal{O}(t^{15})$.}
    \label{fig:v2}
\end{figure}

In Fig.~\ref{fig:v2} we show our results for the development of elliptic flow $v_2(t)$ for two values of the Knudsen number.
The outcome of the simulations is shown in green, where the error bar at $t=0$ is the statistical error due to finite particle number in the initial state of the simulation, as described in Sec.~\ref{subsec:numerical_simulations}.
The blue resp.\ orange curve shows $v_2(t)$ as obtained within the analytical approach starting from the Taylor expansion~\eqref{FullExpansion}, including terms up to order $t^{15}$ but keeping only the linear order in $\sigma$ resp.\ terms up to order $\sigma^3$. 
In Ref.~\cite{Borrell:2021cmh} we showed that the contributions at ${\cal O}(\sigma)$ show up at odd orders starting from $t^3$, those at ${\cal O}(\sigma^2)$ start at order $t^4$ and show up at even order in $t$, while ${\cal O}(\sigma^3)$-terms again show up in odd powers of $t$ from order $t^5$ on. 

For the calculations at the largest Knudsen number (upper panel of Fig.~\ref{fig:v2}) we find an nice agreement between the numerical and analytical results at order $\sigma$ up to $t/R\approx 0.4$.
The agreement extends even further at $\mathcal{O}(\sigma^3)$, where including these extra terms seems to slow down the growth of $v_2(t)$, such that at $t/R = 1$ the two curves differ by less than 10\% --- which is by far not obvious for a Taylor expansion in $t/R$. 

For the five times smaller Knudsen number ${\rm Kn}=5$ (lower panel), the importance of including higher powers of $\sigma$ is even more striking, since it yields a pretty good agreement between the analytical calculations and the simulation up to $t/R\approx 0.6$.

Overall the inclusion of higher orders in the cross section has a major impact on the numerical comparison. 
In particular, it is clear that considering enough powers in $t$ and in $\sigma$ allows to reproduce the behavior that was quantified by a Kn-dependent scaling exponent $\gamma_2$ in Sect.~\ref{subsec:flow_coefficients_Kn}.
Note that further orders in time will indeed improve the curve at late times but not visibly at early times and including those terms is extremely costly computationally.
Similarly higher powers in the cross section would guarantee a better result for smaller Knudsen numbers but we are again limited by the computational power. 

Turning now to triangular flow $v_3$, one first finds~\cite{Bachmann:2022cls} that numerical simulations with the $2\to 0$ model differ more than those with the $2\to 2$ collision kernel, especially at larger times ($t\gtrsim R$).
Looking back at Fig.~\ref{fig:vn_vs_Kn}, we see that at ${\rm Kn} \approx 5$ or 25 the $v_3$ signal at the end of the evolution is already quite small in $2\to 2$ simulations, and thus it is even smaller in $2\to 0$ calculations stopped at $t=R$.
Accordingly, the noise from numerical fluctuations, in particular the small non-zero $v_3$ in the initial state, starts to play an important role --- especially in the simulations at ${\rm Kn}\approx 25$ which we present below.
As argued in Ref.~\cite{Borrell:2021cmh}, a non-vanishing $v_3$ can lead to a (small) linear rise of $v_3(t)$ at early times, which is indeed what we observe.

In addition, there is an unpleasant feature in the analytical calculations with the $2\to 0$ collision term, namely that they yield vanishing odd flow harmonics at all times, in particular $v_3$, if one considers only terms of $\mathcal{O}(\sigma)$~\cite{Bachmann:2022cls}.
On the other hand, an advantage of the Taylor series approach is that we can go to higher orders in $\sigma$, which we attempted here. 

\begin{figure}[!t]
	\centering
	\includegraphics[width=\columnwidth]{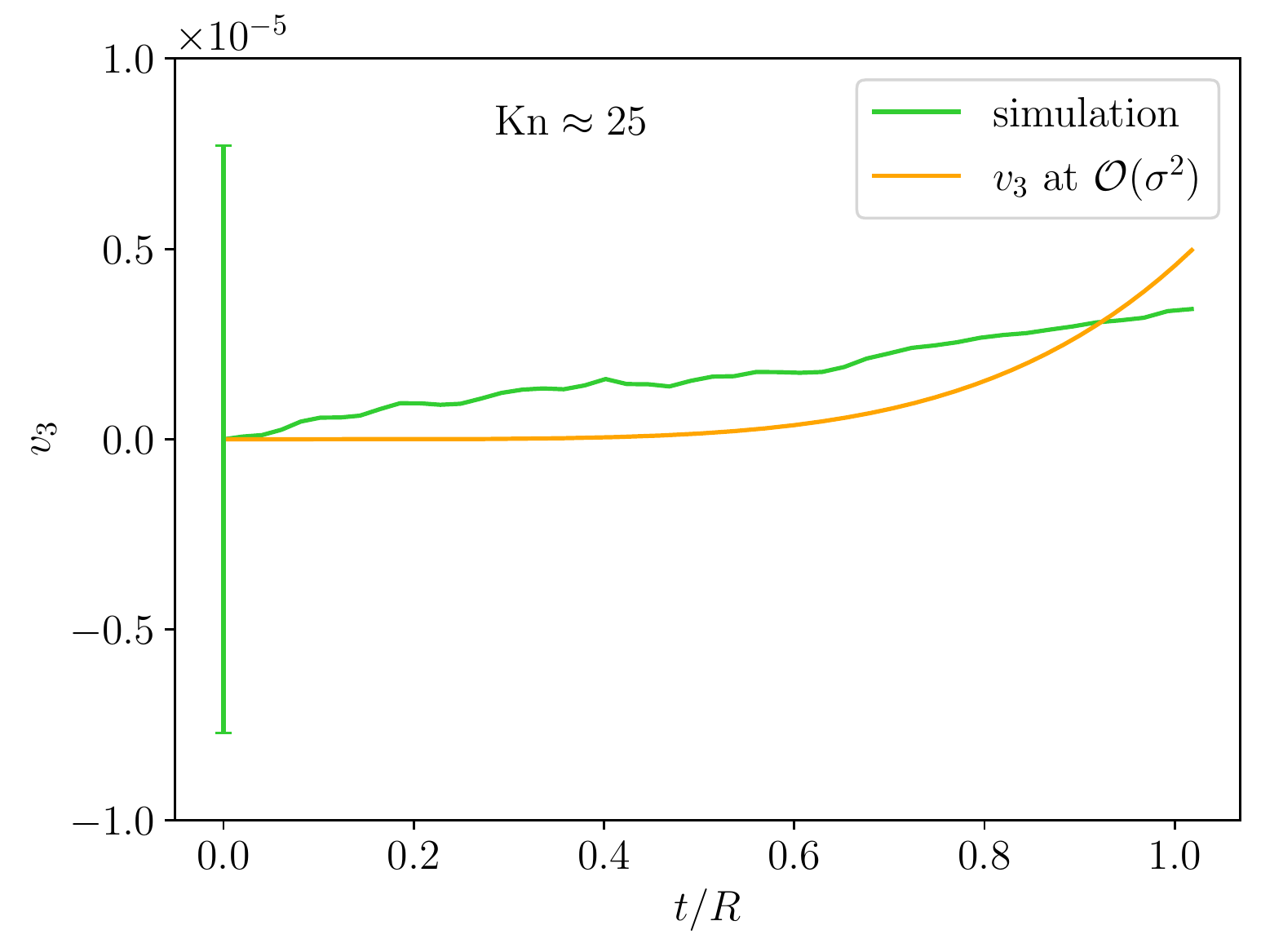}
	\includegraphics[width=\columnwidth]{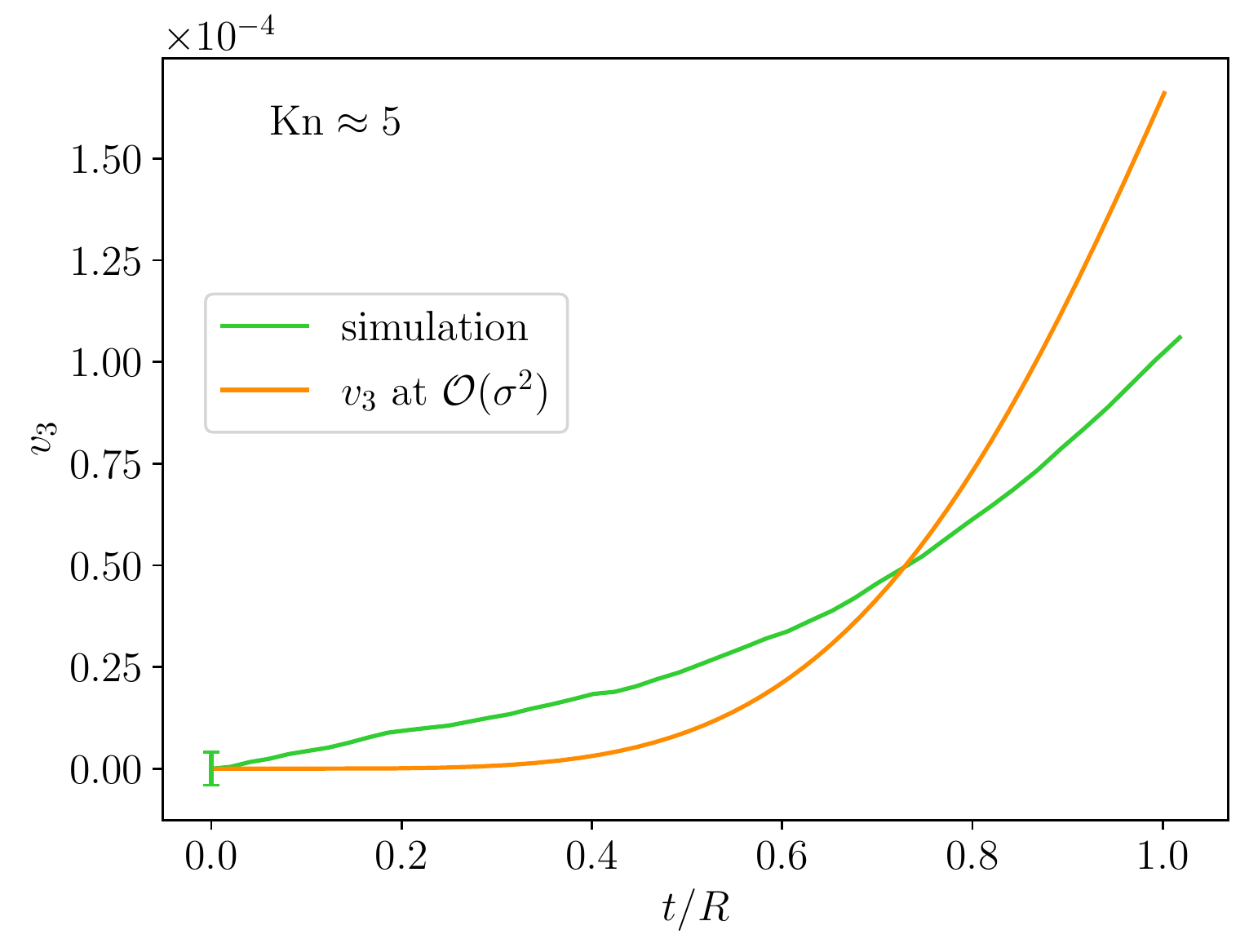}
	\caption{Triangular flow $v_3$ as a function of $t/R$ for $\mathrm{Kn}\approx 25$ (top) and $\mathrm{Kn}\approx 5$ (bottom). Results from the $2\to 0$ collision kernel are shown in green and the analytical results are shown to $\mathcal{O}(\sigma^2)$ (orange).}
	\label{fig:v3}
\end{figure}

With these caveats in mind, we present our results for $v_3(t)$ in Fig.~\ref{fig:v3}, for the same values of the Knudsen number as $N_{\rm resc.}$ and $v_2$.
Here the analytical calculations include terms up to order $t^{11}$ and ${\cal O}(\sigma^2)$.
This means that the leading order at early times is $v_3(t)\propto \sigma^2 (t/R)^5$, i.e.\ a very slow growth.

To remain careful in our statements, let us just notice that the analytical and numerical signals seem to be of the same magnitude. 
This came somewhat as a surprise to us, because it means that in a $2\to 0$ scenario $v_3$ grows as $\sigma^2$, not proportional to $\sigma$. 
In contrast, the simulations with a $2\to 2$ collision kernel clearly give $v_3$ rather proportional to $\sigma$ --- see for instance Fig.~\ref{fig:vn_vs_Kn} and the successful fit with Eq.~\eqref{eq:v3_Kn} which gives $v_3\propto {\rm Kn}^{-1}$ at large Kn.
Nevertheless, this seems to be the message conveyed by our findings in Fig.~\ref{fig:v3}, together with the ``success'' of the analytical approach in describing the numerical results. 

As we shall see in Sect.~\ref{subsec:alternative_flow_observables}, an alternative measure of triangular flow, namely weighted with the particle energy, gives rather different results.

\subsection{Spatial characteristics}
\label{subsec:eccentricities}

Let us now discuss the early-time dependence of various spatial characteristics, for instance the eccentricities $\varepsilon_n^{\bf x}$ or the mean square radius of the expanding system.
A significant difference with anisotropic flow is that most of these characteristics evolve in a free streaming system, in the absence of rescatterings. 
Indeed, it was even found in Ref.~\cite{Borrell:2021cmh} that the change in the geometrical ``observables'' caused by particle rescatterings is at times subleading with respect to the free evolution. 
Accordingly, we first look at collisionless systems, and check that we can describe them in our analytical approach, i.e.\ taking into account only the term $f_{\mathrm{f.s.}}$ in Eq.~\eqref{FullExpansion}, before we turn to systems with (few) collisions. 
Throughout this subsection we come back to simulations with the $2\to 2$ collision kernel, which makes no difference in the free streaming case but is rather crucial when collisions are allowed. 

An important class of geometrical quantities, which are correlated with anisotropic flow via rescatterings, are the spatial eccentricities. 
Here we consider both eccentricities computed via Eq.~\eqref{eccentricity} with particle-number weighting
\begin{equation}
\label{eccentricity_N}
\varepsilon_n^{\bf x}(t)\equiv 
  -\frac{\displaystyle\int\!\! f(t,{\bf x},{\bf p})\,r^n{\rm e}^{{\rm i}n\theta}\,{\rm d}^2{\bf x}\,{\rm d}^2{\bf p}}{\displaystyle\int\!\! f(t,{\bf x},{\bf p})\,r^n\,{\rm d}^2{\bf x}\,{\rm d}^2{\bf p}},
\end{equation}
where we used the fact that for our special case of initial geometry~\eqref{eq:initial_distribution} the reaction-plane angle $\Phi_n$ is zero, and energy-weighted eccentricities $\varepsilon_n^{{\rm x},E}$
\begin{equation}
\label{eccentricity_E}
\varepsilon_n^{{\bf x},E}(t)\equiv 
  -\frac{\displaystyle\int\!\! E\, f(t,{\bf x},{\bf p})\,r^n{\rm e}^{{\rm i}n\theta}\,{\rm d}^2{\bf x}\,{\rm d}^2{\bf p}}{\displaystyle\int\!\! E\, f(t,{\bf x},{\bf p})\,r^n\,{\rm d}^2{\bf x}\,{\rm d}^2{\bf p}},
\end{equation}
with $E\equiv |{\bf p}|$ the energy of a particle with momentum ${\bf p}$.
The behavior of these eccentricities in a free streaming system is readily computed~\cite{Borrell:2021cmh}:
\begin{align}
\label{eq:en_fs}
\varepsilon_n(t) \simeq \frac{\varepsilon_n(0)}{1+a_n(t/R)^2}, 
\end{align}
where the approximate equality is actually exact in the case $n=2$, while for $n\neq 2$ it holds up to terms of order $(t/R)^4$ and possibly higher (if $n\neq 4$).
For the initial condition(s) we consider in this paper, see Sect.~\ref{subsec:distribution_function}, the coefficient $a_2$ equals $1/2$ if a particle number weight is used, i.e.\ for $\varepsilon_2^{\bf x}(t)$, while it is $3/4$ for $\varepsilon_2^{{\bf x},E}(t)$ with energy weight.
In the third harmonic, $a_3$ takes the respective values $3/4$ for $\varepsilon_3^{\bf x}(t)$ and $9/8$ for $\varepsilon_3^{{\bf x},E}(t)$.
Note that these numerical values are valid only for the initial distribution~\eqref{InitialDistribution}--\eqref{eq:initial_distribution}, while Eq.~\eqref{eq:en_fs} holds for any initial distribution in the freely streaming case.

\begin{figure}[!t]
	\centering
	\includegraphics[width=\columnwidth]{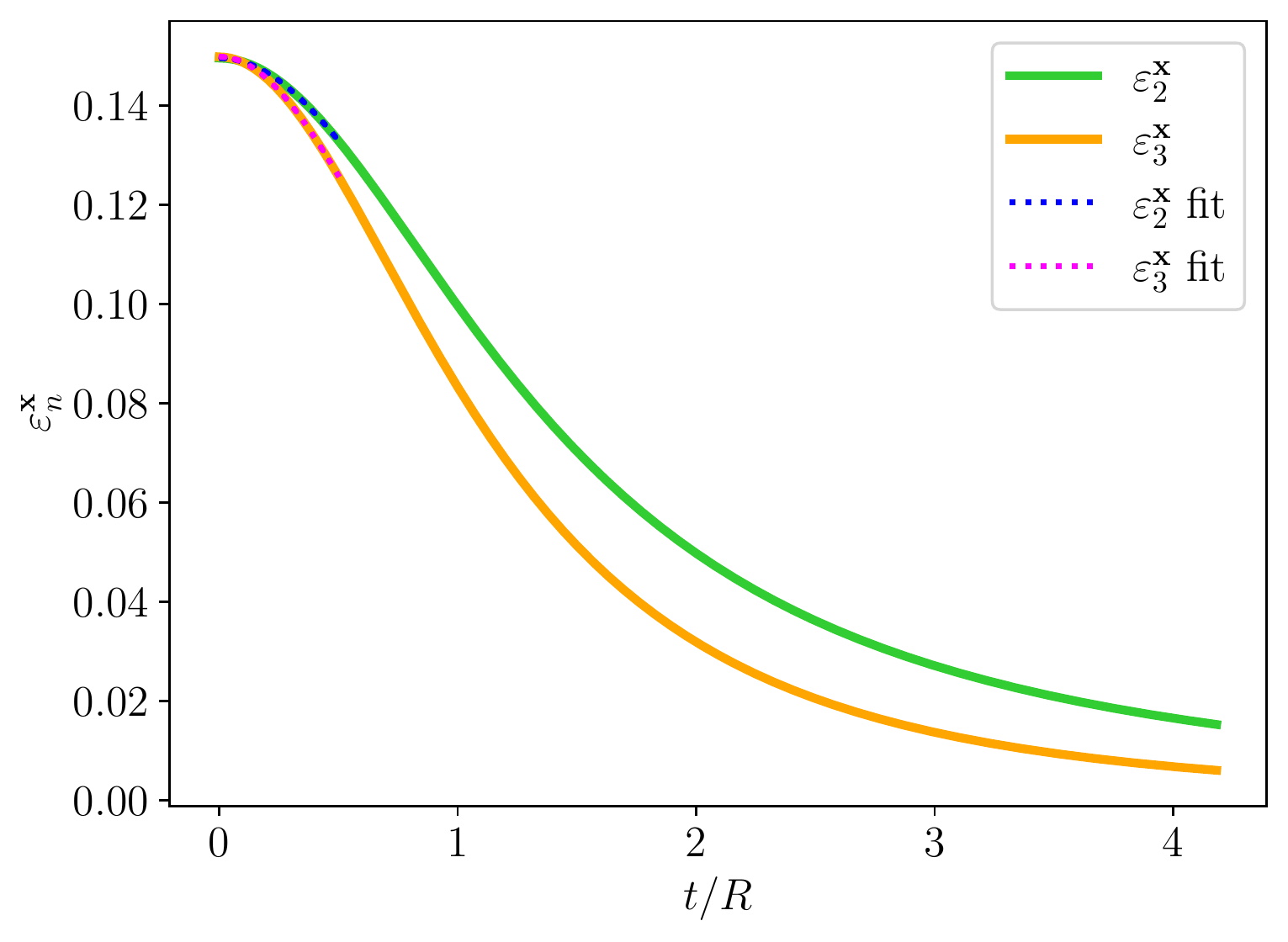}
	\includegraphics[width=\columnwidth]{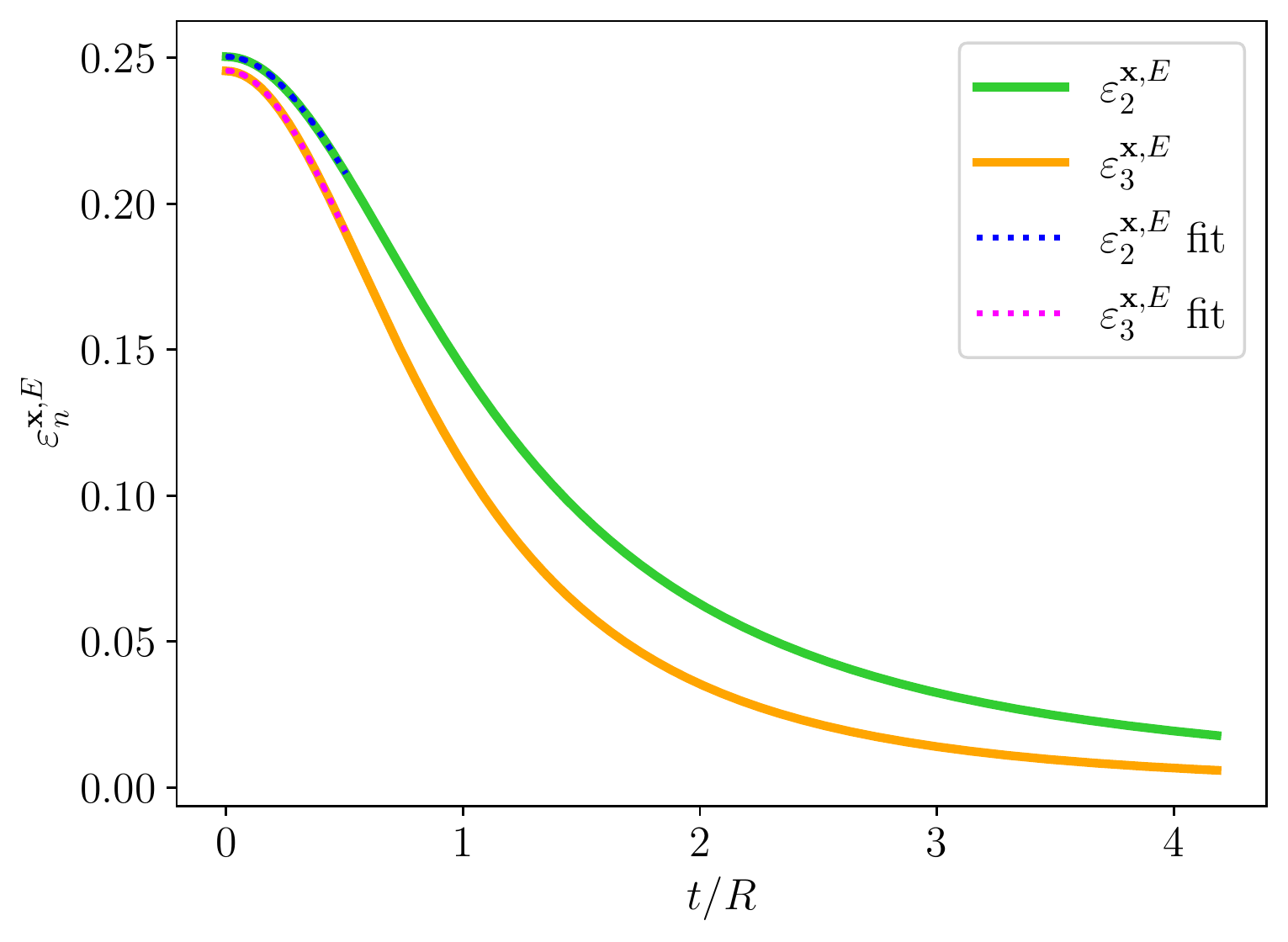}
	\caption{Time evolution of the spatial eccentricities in a collisionless system. The top panel shows particle-number weighted eccentricities $\varepsilon_2^{\bf x}$, $\varepsilon_3^{\bf x}$, the bottom panel eccentricities $\varepsilon_2^{{\bf x},E}$, $\varepsilon_3^{{\bf x},E}$ with energy weighting. The dotted curves are fits with Eq.~\eqref{eq:en_fs} and the parameters given in Table~\ref{tab:en_fs_fit}.}
	\label{fig:en_fs}
\end{figure}

For our numerical calculations, we start with an input value of $\varepsilon_n^{\bf x}=0.15$ in the initial state.
Due to the finite number of particles, which is $N_\mathrm{p} = 2\times 10^5$ in this subsection, this value is not reached exactly in the simulations.\footnote{To minimize fluctuations, we have fixed the initial positions for $N_{\rm iter.} = 3.2\times 10^4$ iterations and resampled the momenta only.}
In addition, the energy-weighted initial eccentricities $\varepsilon_2^{{\bf x},E}$ and $\varepsilon_3^{{\bf x},E}$ can differ from each other, since the momentum (and thus energy) distribution depends on the position. 
Figure~\ref{fig:en_fs} shows the time dependence of the eccentricities in free-streaming.
The full curves showing the free-streaming evolution can be nicely fitted with Eq.~\eqref{eq:en_fs} resulting in the fit parameters of Tab.~\ref{tab:en_fs_fit}, where we fitted the numerical simulations up to $t/R=0.5$.
\begin{table}[!t]
    \centering
    \caption{\label{tab:en_fs_fit}Parameters of the fit function Eq.~\eqref{eq:en_fs} for $\varepsilon_n$ with particle-number and energy weight.}
    \medskip
    
    \begin{tabular}{ccccc}
        \hline\hline
        weight & $\varepsilon_2(0)$ & $a_2$ & $\varepsilon_3(0)$ & $a_3$ \\
        \hline
        particle number & $0.1496$ & $0.5005$ & $0.1499$ & $0.7626$ \\
        energy & $0.250$  & $0.745$ & $0.246$  & $1.141$ \\
        \hline\hline
    \end{tabular}
\end{table}
For particle-number and energy weighting, and for both $n=2$ and $n=3$, the fitted $\varepsilon_n(0)$ coincide with the initial values computed directly at the beginning of the simulations. 
In turn, the fitted values of $a_n$ are in good agreement with the analytical prediction: as could be anticipated, the agreement is better for $n=2$, in which case Eq.~\eqref{eq:en_fs} is exact, than for $n=3$, in which case the denominator contains higher (even) powers of $t$. 
Note that the energy-weighted eccentricities $\varepsilon_n^{{\bf x},E}$ decrease slightly faster than their particle-number weighted counterparts, which is reflected in the larger values of $a_n$.

\begin{figure}[!t]
	\centering
	\includegraphics[width=\columnwidth]{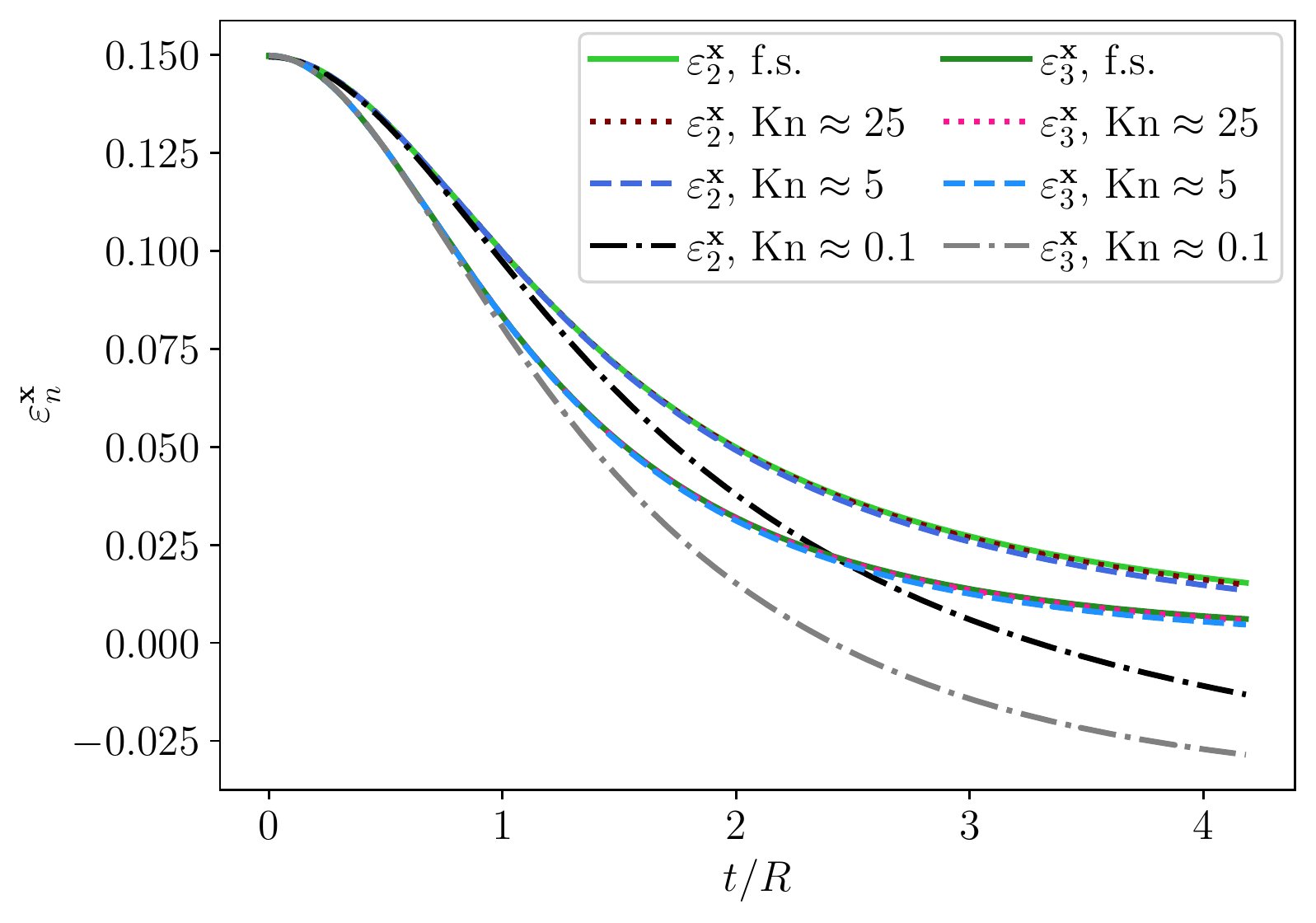}
	\includegraphics[width=\columnwidth]{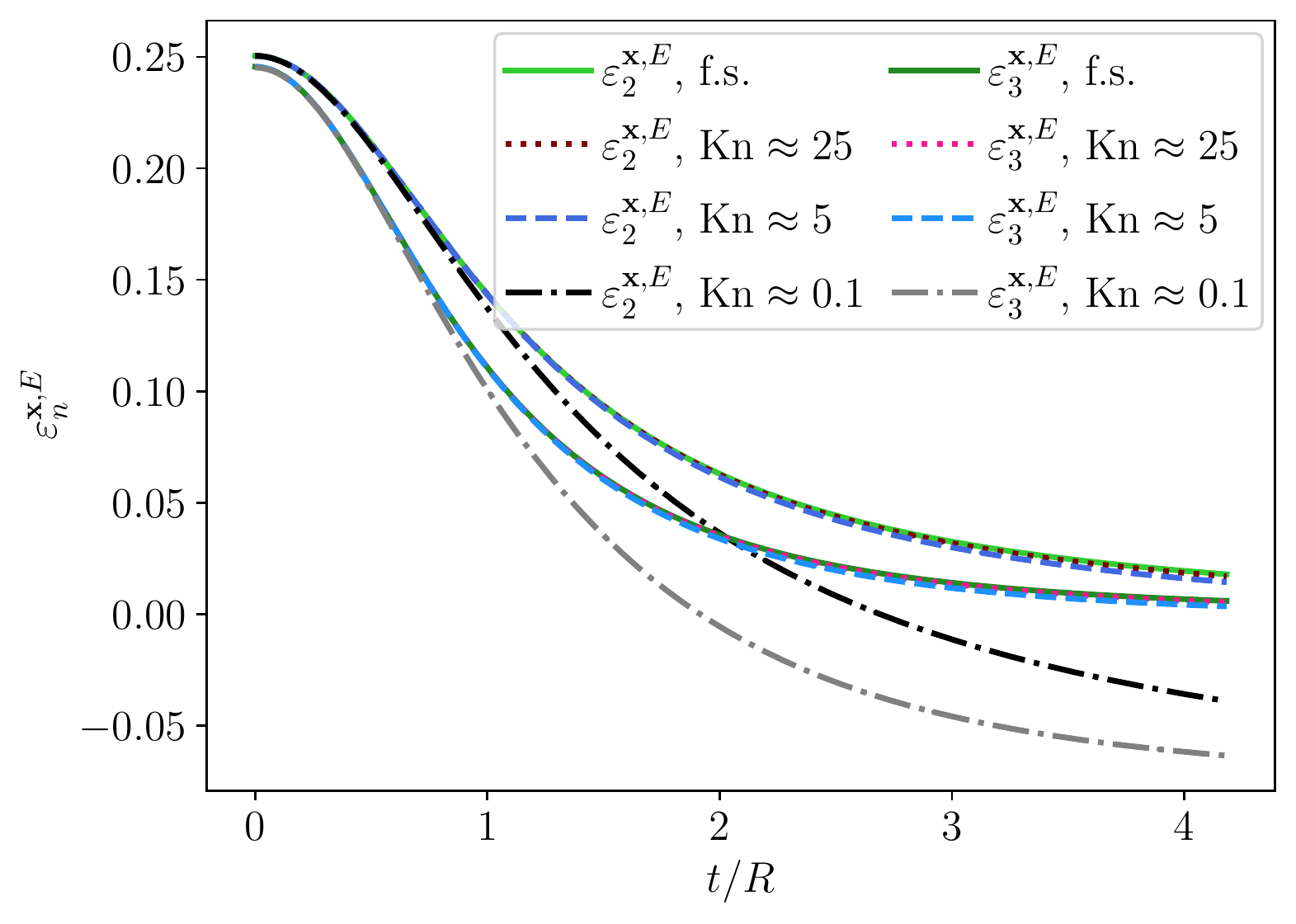}
	\caption{Time evolution of the spatial eccentricities in a collisionless (full lines) and in an interacting system with few rescatterings (dotted and dashed lines). The top panel shows particle-number weighted eccentricities $\varepsilon_2^{\bf x}$, $\varepsilon_3^{\bf x}$, the bottom panel eccentricities $\varepsilon_2^{{\bf x},E}$, $\varepsilon_3^{{\bf x},E}$ with energy weighting.
	For comparison, we also show the behavior in a system in the fluid-dynamical regime (dotted-dashed lines).}
	\label{fig:en_collisions}
\end{figure}

In the free-streaming case there is no transfer of spatial asymmetry into a momentum anisotropy and thus no buildup of anisotropic flow.
That is the reason why the eccentricity decreases most slowly in the collisionless case: introducing rescatterings in the system will lead to a faster decrease.
With collisions it is even possible to change the sign of the eccentricity --- as observed in fluid-dynamical simulations~\cite{Kolb:2002cq}, while for the free-streaming case this does not happen.
However, at large Knudsen numbers (${\rm Kn}\approx 25$ or 5) the change in the evolution of eccentricities due to rescatterings is barely visible (see Fig.~\ref{fig:en_collisions}), even at large times.
For reference, we also include the behavior of the eccentricities for a system close to the fluid-dynamical regime, namely at ${\rm Kn}\approx 0.1$. 
One sees that sizable deviations from the collisionless behavior only appear for $t\gtrsim R$. 
This is consistent with the fact that it takes about a similar time for anisotropic flow to acquire a sizable value across the whole system, and thus to affect the evolution of geometry.

To pinpoint the influence of rescatterings, let us investigate the behaviors of the numerator and denominator of the eccentricity separately.
Here we only discuss results relevant for $\varepsilon_2^{\bf x}$ and $\varepsilon_2^{{\bf x},E}$, since the behaviors in the third harmonic (shown in Appendix~\ref{appendix:epsilon3}) are similar.
To see the effect of collisions more easily, we will subtract the free-streaming behaviors.

\begin{figure}[!t]
	\centering
	\includegraphics[width=\columnwidth]{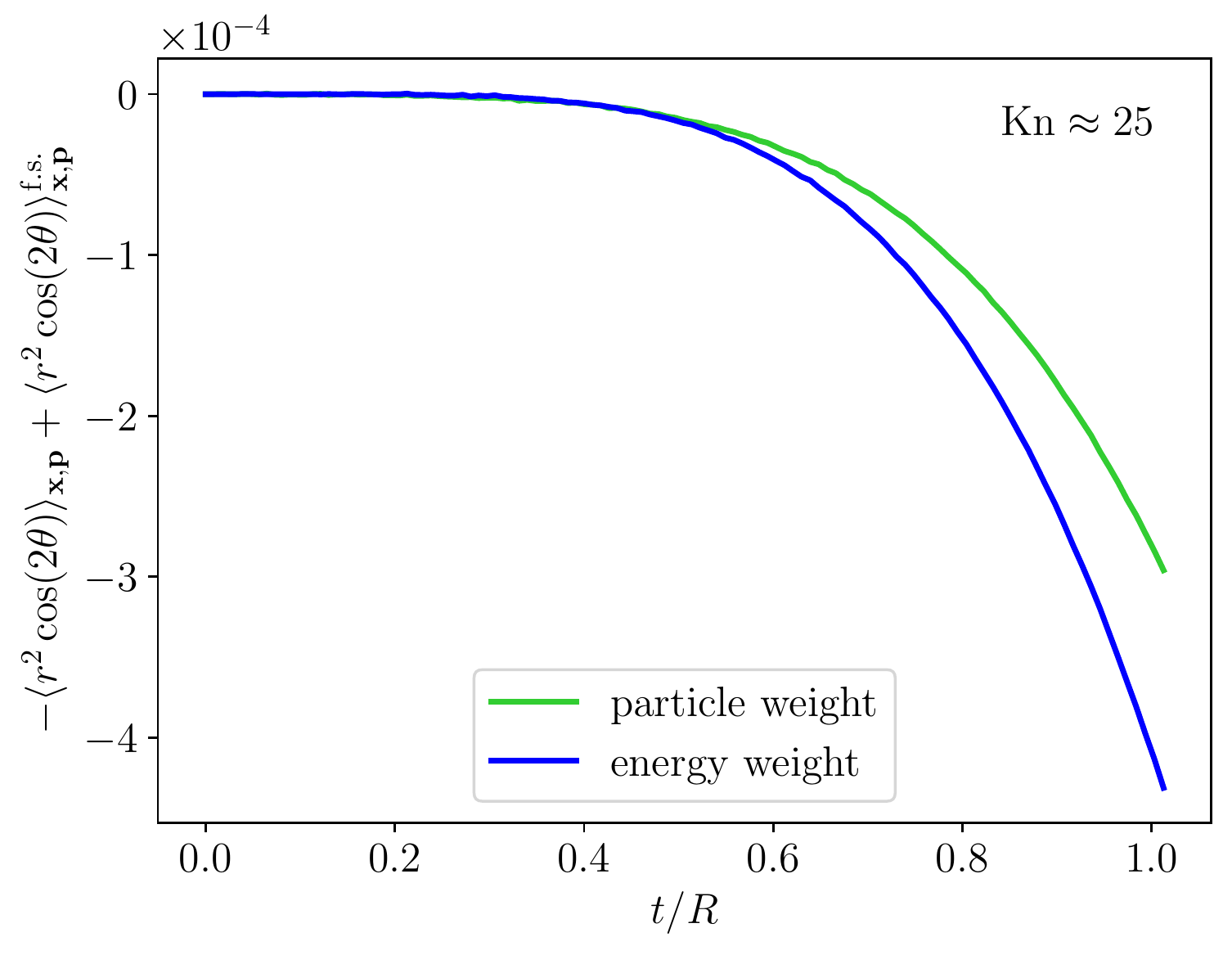}
	\includegraphics[width=\columnwidth]{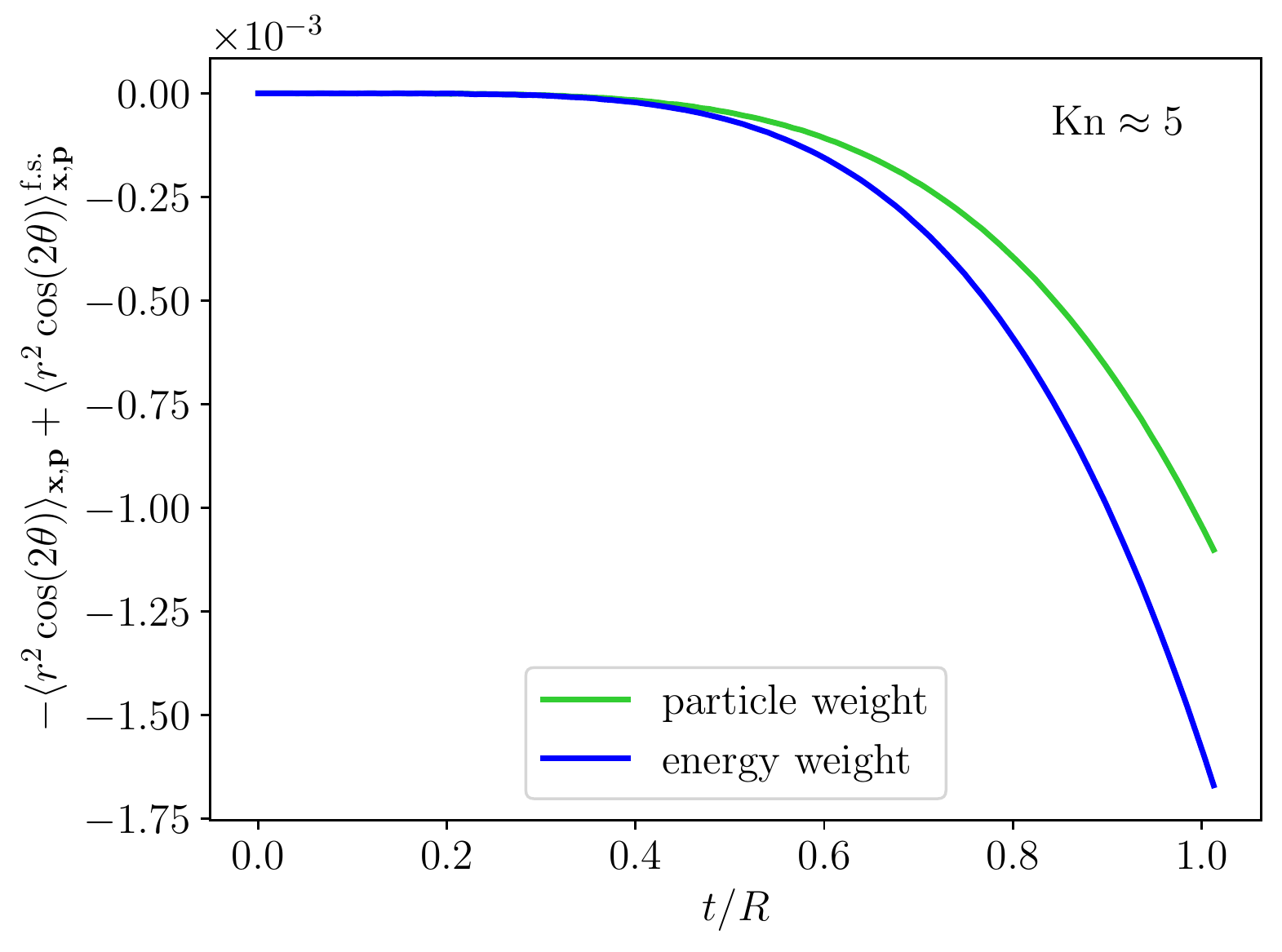}
	\caption{Time evolution of the departure of $-\langle r^2\cos(2\theta)\rangle_{{\bf x},{\bf p}}$ with particle-number (green) and energy (blue) weighting from its free streaming value, for simulations with $\mathrm{Kn}\approx 25$ (top panel) or $\mathrm{Kn}\approx 5$ (bottom panel).}
	\label{fig:rsquaredcos2theta}
\end{figure}

Let us start with the numerator of the eccentricity, i.e.\ (minus) the average value of $r^2\cos(2\theta)$, where the averaging can be done with particle-number or energy weighting.
This is actually the simpler case, since in the free streaming case --- and in the absence of initial anisotropic flow, as we assume throughout this subsection --- that quantity remains constant over time: any evolution is thus due to rescatterings. 
We show this departure from the free-streaming value in Fig.~\ref{fig:rsquaredcos2theta} for simulations with Knudsen number ${\rm Kn} \approx 25$ (top panel) and ${\rm Kn} \approx 5$ (bottom panel).
To obtain dimensionless quantities, we scaled the curves by the initial value of the corresponding average of $r^2\cos(2\theta)$: each curve thus represents the relative change in the average value due to rescatterings. 
One sees that this relative change is of order a few $10^{-4}$ for the system with ${\rm Kn} \approx 25$, and of order $10^{-3}$ when ${\rm Kn} \approx 5$. 
The latter signal is thus about 5 times larger than the former, i.e.\ seems to scale with ${\rm Kn}^{-1}$ or equivalently $\sigma$ or the number of rescatterings in the system. 

To quantify the behaviors displayed in Fig.~\ref{fig:rsquaredcos2theta}, we tried to fit the curves with a simple power-law ansatz
\begin{equation}
-\!\langle r^2\cos(2\theta)\rangle_{{\bf x},\bf p}(t) + \langle r^2\cos(2\theta)\rangle_{{\bf x},\bf p}^{\rm f.s.} \propto \left(\frac{t}{R}\right)^{\!\!\chi}.
    \label{eq:r2cos2thetafit}
\end{equation}
As in Sect.~\ref{subsec:flow_coefficients_Kn}, we perform the fit over a time interval from $t=0$ to some final $t/R$ in the range 0.35--0.7, letting the end point of the interval vary.
The fit results for the exponents lie in the range $\chi = 3.5$--5.5. 
This is a rather slow growth, and we know from our fitting of the early-time behavior of $v_3(t)$ in Sect.~\ref{subsec:flow_coefficients_Kn} that it probably points at an exponent $\chi\geq 4$, but we do not want to make any strong claim. 
Indeed, the signal of Fig.~\ref{fig:rsquaredcos2theta} is one to two orders of magnitude smaller than that of $v_3(t)$. 

With our analytical approach, one finds invoking general principles --- namely particle-number or energy conservation, depending on the weight of the average --- that the contributions from the Taylor expansion~\eqref{FullExpansion} at order $t^2$ and below vanish, leaving $\chi\geq 3$~\cite{Borrell:2021cmh}. 
We can further show that the contributions from the loss term of the $2\to 2$ collision kernel vanish at any order, invoking parity arguments in momentum space. 
Unfortunately, we were unable to compute the contribution from the gain term of the collision integral, so we cannot give a more accurate prediction than the inequality $\chi\geq 3$, in agreement with the fit values that we find.

Turning now to the denominator of $\varepsilon_2^{\bf x}$, i.e.\ the mean square radius, it differs from the numerator in that it is evolving in a collisionless system:
\begin{equation}
\langle r^2\rangle_{{\bf x},{\bf p}}^{\rm f.s.}(t) = 
  \langle r^2\rangle_{{\bf x},{\bf p}}(0) + 
  \langle v^2\rangle_{{\bf x},{\bf p}}(0)\,t^2,
\end{equation}
where $v^2$ denotes the squared (transverse) velocity of the particles --- here simply equal to the squared velocity of light since we consider massless particles.

\begin{figure}[!t]
	\centering
	\includegraphics[width=\columnwidth]{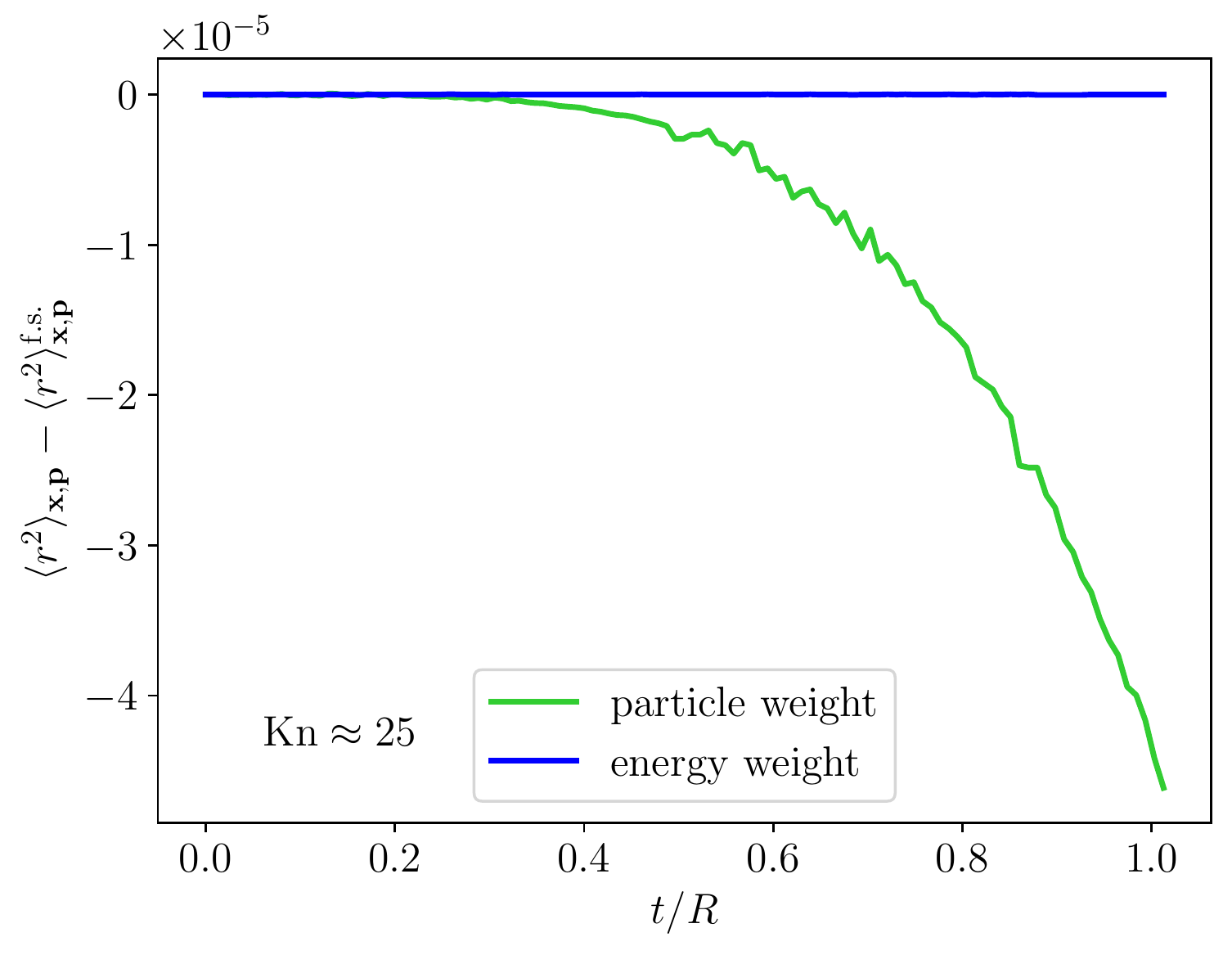}
	\includegraphics[width=\columnwidth]{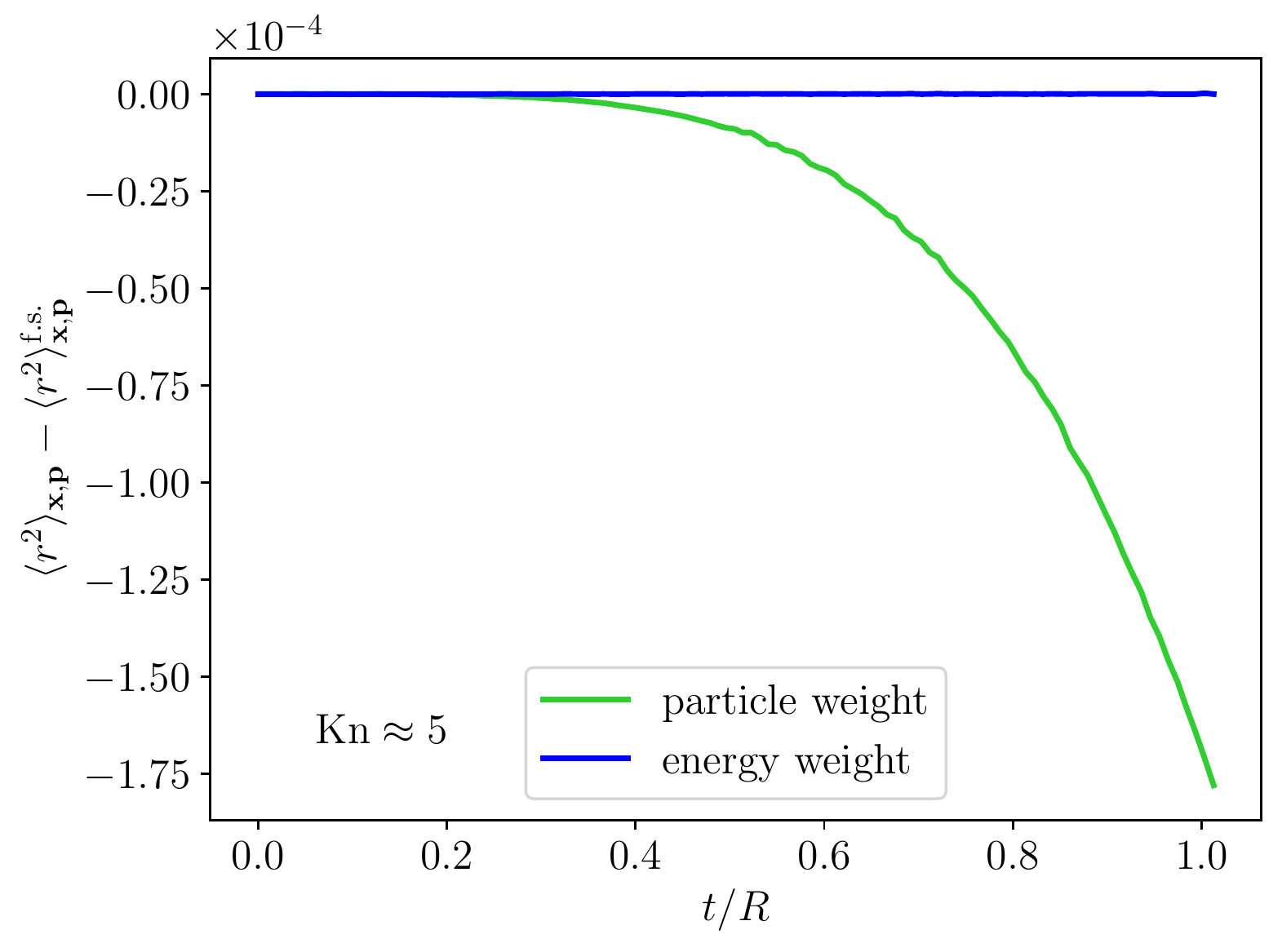}
	\caption{Time evolution of the departure of $\langle r^2\rangle_{{\bf x},{\bf p}}$ with particle-number (green) or energy (blue) weighting from its free streaming behavior, for simulations with $\mathrm{Kn}\approx 25$ (top panel) or $\mathrm{Kn}\approx 5$ (bottom panel).}
	\label{fig:rsquared}
\end{figure}

We show in Fig.~\ref{fig:rsquared} the departure $\langle r^2\rangle_{{\bf x},{\bf p}}(t)-\langle r^2\rangle_{{\bf x},{\bf p}}^{\rm f.s.}(t)$ of the mean square radius in a interacting system from this free-streaming evolution, scaled by the initial value of $\langle r^2\rangle_{{\bf x},{\bf p}}$. 
The departure is negative, which means that rescatterings slow down the system expansion: 
this seems intuitive, since collisions cannot accelerate the outwards motion of particles already traveling at the speed of light.
As in the case of $\langle r^2\cos(2\theta)\rangle_{{\bf x},{\bf p}}$, one sees that the relative deviation of the mean square radius computed with particle weight from its free-streaming behavior (green curves) changes by about a factor of 5 from ${\rm Kn}\approx 25$ to ${\rm Kn}\approx 5$, i.e.\ it seems to scale linearly with the number of rescatterings. 
A power-law fit of these curves
\begin{align}
\langle r^2\rangle_{{\bf x},{\bf p}} - \langle r^2\rangle_{{\bf x},{\bf p}}^\mathrm{f.s.} \propto 
  \left(\frac{t}{R}\right)^{\!\!\chi^\prime}
\label{eq:r2fit}
\end{align}
yield exponents in the range $\chi'=4.1$--4.5, depending on the end point of the time interval chosen for the fit. 
In the analytical approach we found $\chi'\geq 3$~\cite{Borrell:2021cmh}, consistent with the numerical finding. 

Quite strikingly, the blue curves in Fig.~\ref{fig:rsquared} representing the energy-weighted mean square radius stay constant. 
That is, rescatterings in the system do not modify the evolution of the energy-weighted $\langle r^2\rangle_{{\bf x},{\bf p}}$, i.e.\ of the denominator of $\varepsilon_2^{{\bf x},E}(t)$. 
We detail in Appendix~\ref{appendix:<Er^2>} how this can actually be shown within our analytical approach~\cite{Borrell:2021cmh}, mostly invoking energy and momentum conservation --- one has however to beware that it holds only for a two-dimensional system of massless particles.
This absence of departure from the free-streaming behavior is specific to the mean square radius: as shown in  Fig.~\ref{fig:rcubed} in Appendix~\ref{appendix:epsilon3}, the evolution of the energy-weighted average of the cubed radius does depend on the amount of rescatterings in the system. 
Yet the qualitative differences between particle-number weighted and energy-weighted quantities like the moments $\langle r^2\rangle_{{\bf x},{\bf p}}$, $\langle r^3\rangle_{{\bf x},{\bf p}}$ (Figs.~\ref{fig:rsquared} and \ref{fig:rcubed}) show that the transport of energy density in the system does not match one-to-one that of particle-number density.

\subsection{Alternative measures of anisotropic flow}
\label{subsec:alternative_flow_observables}

Motivated by the observation at the end of the previous subsections, we now look at alternative measures of anisotropic collective flow. 
We shall compute the latter in both numerical simulations and analytical calculations and check if their early-time behaviors agree with the results obtained for $v_n(t)$ in Sect.~\ref{subsec:flow_coefficients_Kn} and \ref{subsec:flow_coefficients}.

One possibility, which was adopted in a number of recent  kinetic theory studies~\cite{Kurkela:2018ygx,Kurkela:2019kip,Kurkela:2020wwb,Kurkela:2021ctp,Ambrus:2021fej}, is to consider the Fourier harmonics of the transverse energy distribution. 
Using the terminology of Sect.~\ref{subsec:eccentricities}, these are effectively ``energy-weighted anisotropic flow coefficients''
\begin{equation}
	\label{v_n^E_def}
v_n^{E\,}(t) \equiv 
  \frac{\displaystyle\int\!E_{}f(t,{\bf x},{\bf p})\,\cos[n(\phi_{\bf p}-\Psi_n^E)]\,{\rm d}^2{\bf x}\,{\rm d}^2{\bf p}}%
    {\displaystyle\int\!E_{}f(t,{\bf x},{\bf p})\,{\rm d}^2{\bf x}\,{\rm d}^2{\bf p}},
\end{equation}
with $\Psi_n^E$ the $n$-th harmonic event plane.
In contrast, the traditional coefficients~\eqref{vn_def} are ``particle-number weighted'' flow harmonics.
In the following $\Psi_n^E=0$, consistent with the orientation of the initial-state symmetry planes $\Phi_n=0$.

For massless particles, as we consider throughout the paper, with $p_z=0$, the energy-weighted elliptic flow $v_2^E$ actually coincides with a definition that has been used in fluid-dynamical simulations~\cite{Ollitrault:1992bk,Kolb:2000sd},%
\footnote{Note that there is a misprint in the definition of $\varepsilon_2^{\bf p}$ in Eq.~(3.2) of Ref.~\cite{Kolb:2000sd}, as mentioned in Ref.~\cite{Teaney:2009qa}.}
namely (in our two-dimensional setup)
\begin{align}
\varepsilon_2^{\bf p}\equiv\frac{\displaystyle\int\!\big[T^{xx}({\bf x})-T^{yy}({\bf x})\big]_{}{\rm d}^2{\bf x}}%
  {\displaystyle\int\!\big[T^{xx}({\bf x})+T^{yy}({\bf x})\big]_{}{\rm d}^2{\bf x}}.
\label{eq:e2p}
\end{align}
Note that the notation emphasizes the similarity to the geometrical eccentricities. 
Definition~\eqref{eq:e2p} involves two diagonal components $T^{xx}$ and $T^{yy}$ of the energy-momentum tensor
\begin{align}
T^{\mu\nu}({\bf x}) \equiv 
  \int\! p^\mu p^\nu f(t,{\bf x},{\bf p})\;\frac{{\rm d}^2{\bf p}}{E}.
\label{eq:Tmunu}
\end{align}

An advantage of Eq.~\eqref{eq:e2p} is that it can also be implemented for a system that is not described as a collection of particles or a particle density distribution, but rather in terms of its energy momentum tensor.
In particular, there is no need to ``particlize'' the system using a Cooper--Frye-like approach.

\begin{figure}[!t]
	\centering
	\includegraphics[width=\columnwidth]{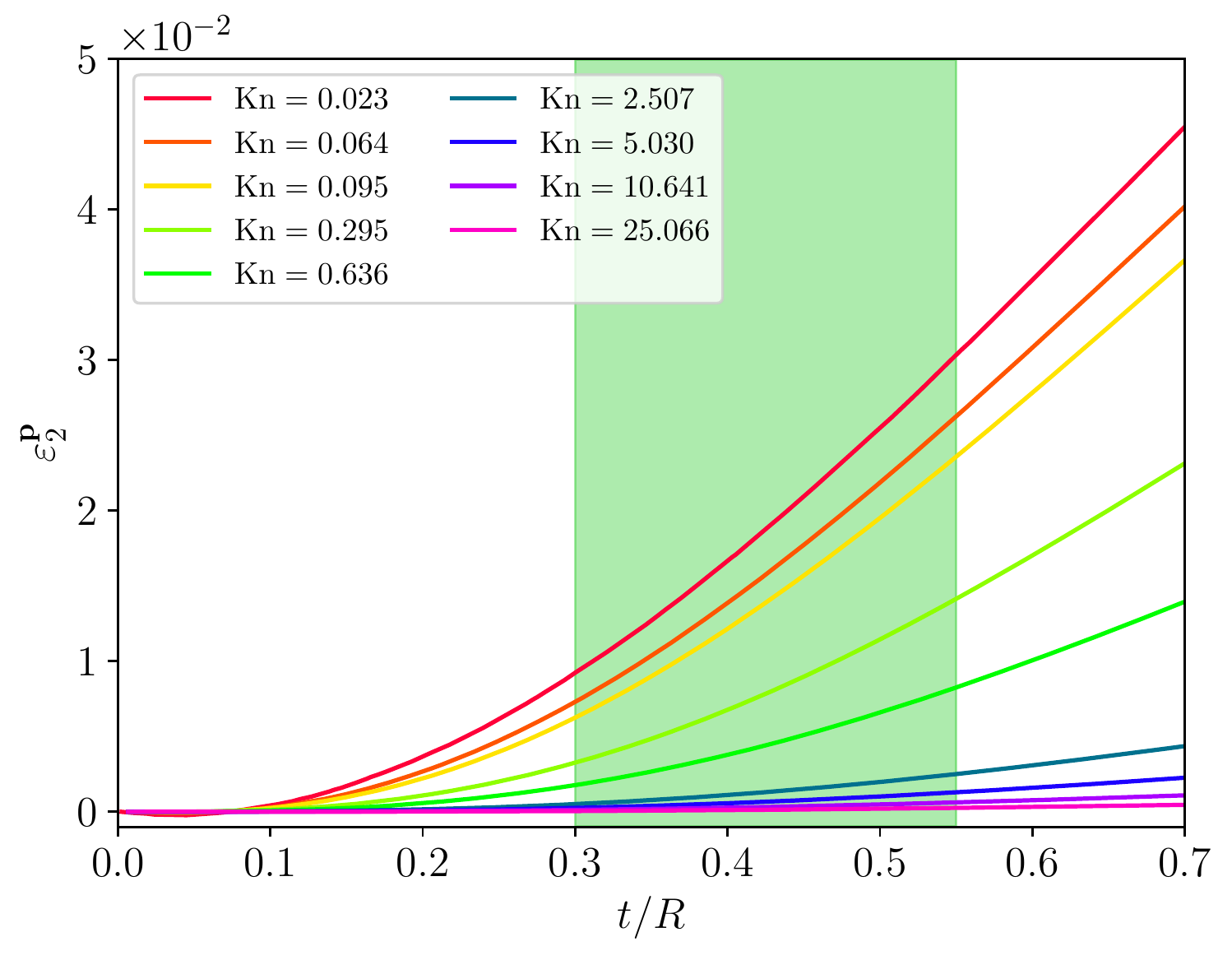}
	\includegraphics[width=\columnwidth]{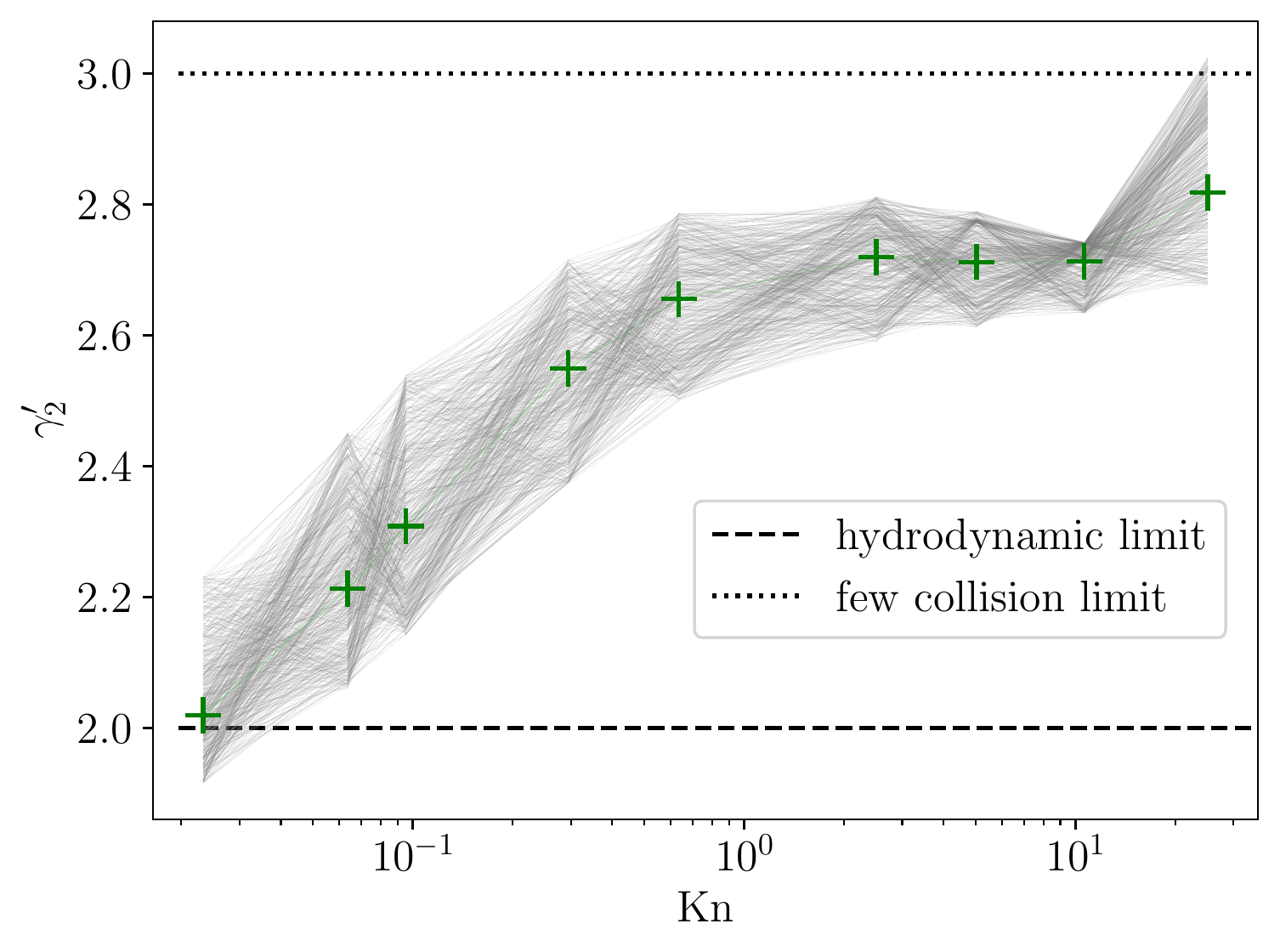}
	\caption{Upper panel: Early time dependence of $\varepsilon_2^{\bf p} = v_2^E$ for various Knudsen numbers. Lower panel: Dependence of the scaling exponent $\gamma'_2$ [Eq.~\eqref{eq:eps2p_time}] on the Knudsen number, using the same color code as in Fig.~\ref{fig:gamma}.}
	\label{fig:eta_2}
\end{figure}

As in Sect.~\ref{subsec:flow_coefficients_Kn}, let us first look at the behavior of $\varepsilon_2^{\bf p}$ across a large range of Knudsen numbers. 
The results of simulations with the $2\to 2$ collision kernel from the few-rescatterings regime (${\rm Kn}\approx 25$) to the fluid-dynamical limit (${\rm Kn}\approx 0.02$) for the onset of $\varepsilon_2^{\bf p}(t) = v_2^E(t)$ is displayed in the top panel of Fig.~\ref{fig:eta_2}.
Similar to $v_2(t)$, we fit the early time development with a power law ansatz
\begin{equation}
\varepsilon_2^{\bf p}\bigg(\frac{t}{R}\bigg) =
   \beta'_2\bigg(\frac{t}{R}\bigg)^{\!\!\gamma'_2},
\label{eq:eps2p_time}
\end{equation}
performing 500 different realizations of the fit over varying time intervals. 
The change with Kn of the resulting scaling exponents $\gamma'_2$ is shown in the bottom panel of Fig.~\ref{fig:eta_2}.
The dependence of $\gamma'_2$ on the Knudsen number closely parallels that of $\gamma_2$ (Fig.~\ref{fig:gamma}), seemingly ranging from 2 in the fluid-dynamical limit --- as found in earlier hydrodynamical simulations --- to roughly 3 in the few-rescatterings regime.
That is, $\varepsilon_2^{\bf p}(t)$ behaves as $v_2(t)$ at early times.

\begin{figure}[!t]
	\centering
	\includegraphics[width=\columnwidth]{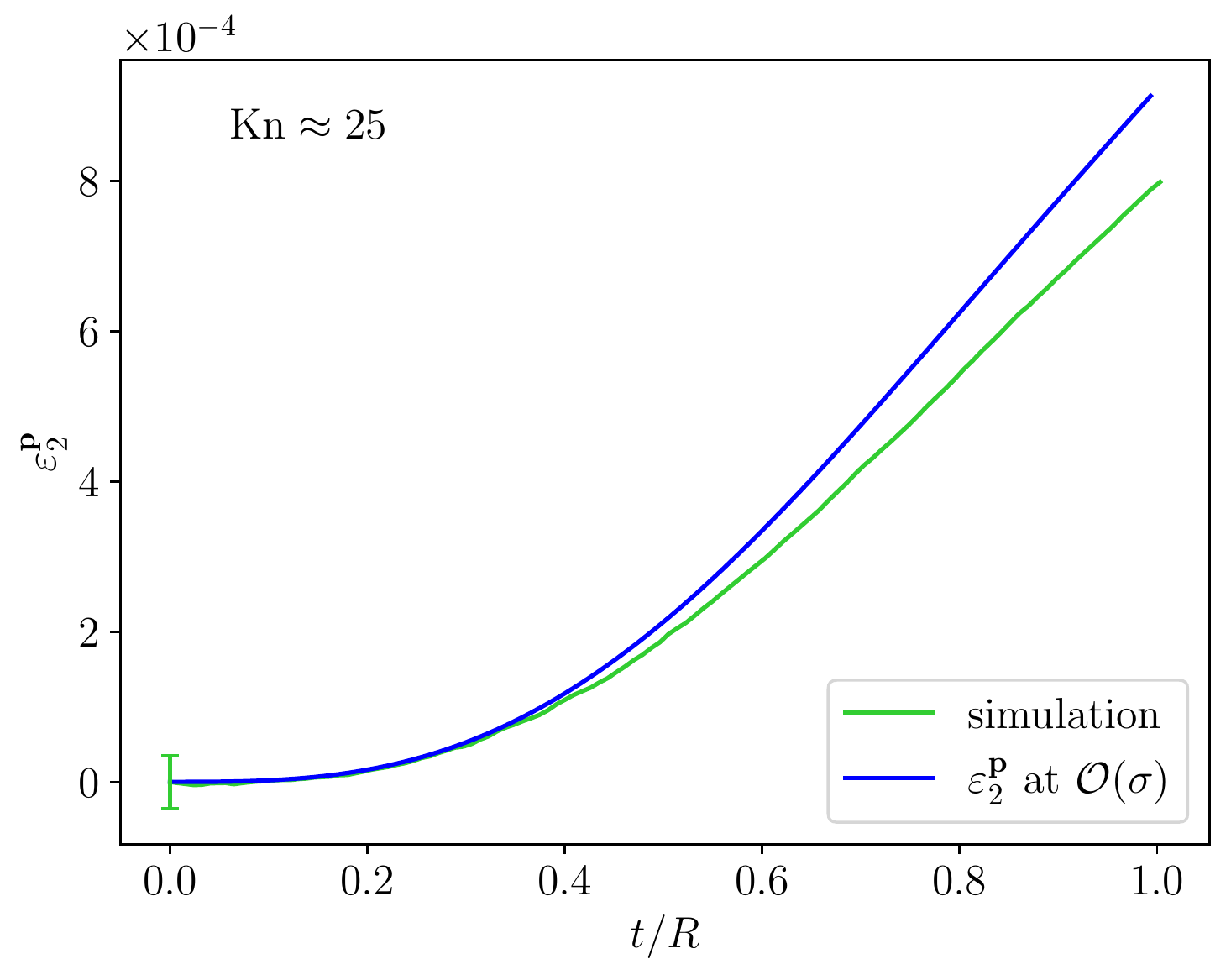}
	\caption{Early time evolution of $\varepsilon_2^{\bf p} = v_2^E$ at $\mathrm{Kn}\approx 25$. The results of the $2\to 0$ simulation are shown in green, the analytical calculation at ${\cal O}(\sigma)$ in blue.}
	\label{fig:eps2p_Kn25_NumAnalyt}
\end{figure}

Using the same arguments as in Ref.~\cite{Borrell:2021cmh} (especially Sect.~IV.1), one finds that in the absence of initial flow, the leading contribution to $\varepsilon_2^{\bf p}(t)$ at early times is
\begin{equation}
\label{eps2p(t)_early-t}
\varepsilon_2^{\bf p}(t) \propto \sigma t^3 + \mathcal{O}(t^4),
\end{equation}
where the term at order $t^4$ is actually of order $\sigma^2$. 
This is similar to the scaling of $v_2(t)$ and is shown as dotted line in the lower panel of Fig.~\ref{fig:eta_2}.
We computed the proportionality coefficient --- as well as terms up to ${\cal O}(t^9)$ at linear order in $\sigma$ --- for the $2\to 0$ collision kernel. 
This is compared to results from numerical simulations, also with the $2\to 0$ kernel, in Fig.~\ref{fig:eps2p_Kn25_NumAnalyt}.
The results from both approaches are in good agreement until $t/R\simeq 0.4$, and we know from our investigations of $v_2(t)$ that we can improve the agreement by including higher orders in $\sigma$ and $t$ in the (already extensive) analytical calculations.

Setting $n=3$ in Eq.~\eqref{v_n^E_def} gives the energy-weighted triangular flow $v_3^E$. 
\begin{figure}[!htb]
	\centering
	\includegraphics[width=\columnwidth]{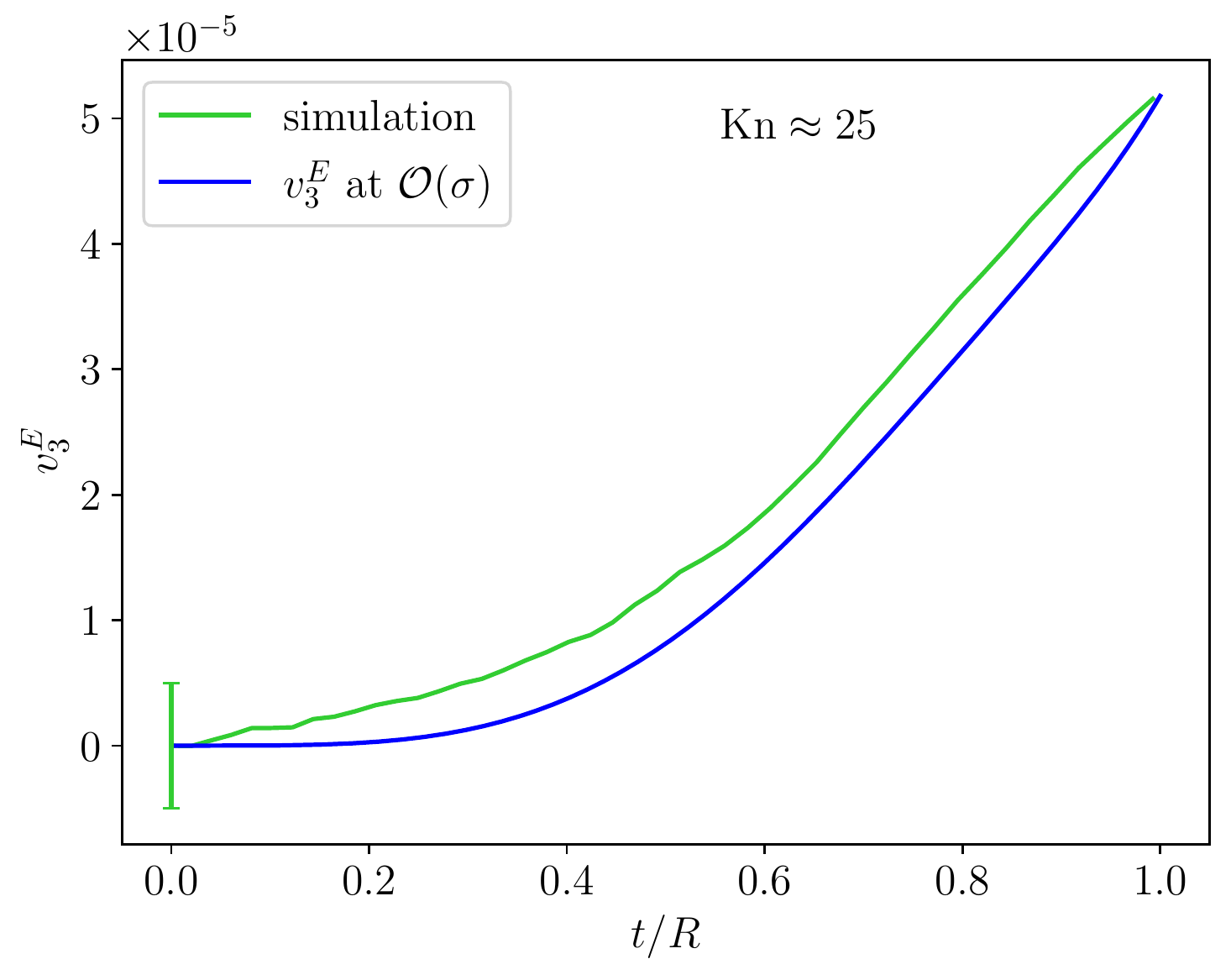}
	\includegraphics[width=\columnwidth]{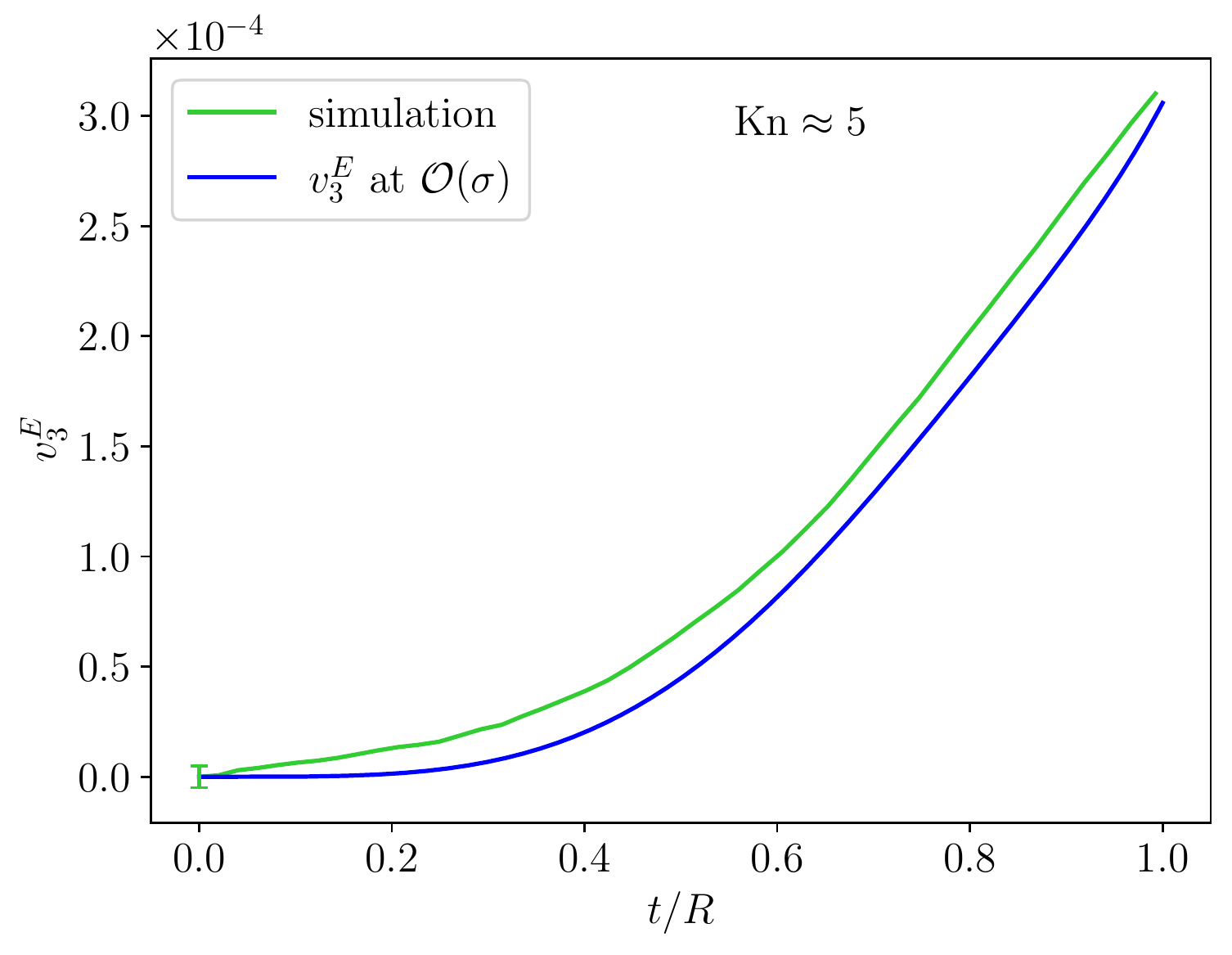}
	\caption{\label{fig:v3_energy}Early time development of energy-weighted triangular flow $v_3^E$ for $\mathrm{Kn}\approx 25$ (top panel) and $\mathrm{Kn}\approx 5$ (bottom panel) from numerical simulations with the $2\to 0$ collision kernel (green) and analytical calculations at $\mathcal{O}(\sigma)$ (blue).}
\end{figure}
We show in Fig.~\ref{fig:v3_energy} results of simulations (green curves) with the $2\to 0$ collision kernel at both $\mathrm{Kn}\approx 25$ (top panel) and $\mathrm{Kn}\approx 5$ (bottom panel). 
Comparing with the particle-number weighted $v_3(t)$ in Fig.~\ref{fig:v3}, the $v_3^E(t)$ signal is significantly larger, and in particular less sensitive to noise in the initial state. 

Figure~\ref{fig:v3_energy} also shows the results of analytical calculations (blue curve), where in contrast to $v_3$ we already find a non-vanishing $v_3^E$ at $\mathcal{O}(\sigma)$ and including terms up to order $t^{10}$ --- while in the same $2\to 0$ scenario $v_3$ was found to scale as ${\cal O}(\sigma^2)$.
The analytical results and those of simulations are actually in quite good agreement for both Knudsen numbers being considered, especially given the relative size of the initial fluctuations of the numerical signal, which yield a visible linear rise at small $t/R$. 
Other differences like the different curvature as $t/R\approx 1$ could certainly be improved by including more orders in $\sigma$ and in $t/R$ in the analytical approach.

As mentioned above, a drawback of the coefficients $v_n$ or $v_n^E$ from the theorist's point of view is that their definition assumes a particle-based model. 
For massless particles with $p_z=0$, $v_2^E$ coincides with the ``momentum space eccentricity'' $\varepsilon_2^{\bf p}$, Eq.~\eqref{eq:e2p}, defined in terms of the energy momentum tensor, but the latter has no straightforward generalization to higher harmonics. 
A possible measure of anisotropic flow at the ``macroscopic'' level --- i.e.\ a priori usable for any microscopic model --- was proposed in Ref.~\cite{Teaney:2010vd} for hydrodynamical calculations, which has the advantage of being generalizable for any harmonic. 
The simplest example is the measure of elliptic flow, via\footnote{We use a different notation from the original one~\cite{Teaney:2010vd} and adapt it to our 2-dimensional setup.}
\begin{align}
\alpha_2^{\bf p} \equiv \frac{\displaystyle\int\!\big[T^{0x}({\bf x})_{}u_x({\bf x}) - T^{0y}({\bf x})_{}u_y({\bf x})\big]\,{\rm d}^2{\bf x}}%
  {\displaystyle\int\!T^{00}({\bf x})_{}u_0({\bf x})\,{\rm d}^2{\bf x}},
\label{eq:alpha2p}
\end{align}
where $u^\mu({\bf x})$ is the system flow velocity.
For a collection of particles of a single species as considered in the present paper, the natural choice for that velocity is
\begin{equation}
u^\mu({\bf x}) =  \frac{N^{\mu}({\bf x})}{\sqrt{N^{\mu}({\bf x})N_{\mu}({\bf x})}},
\label{eq:umu}
\end{equation}
with
\begin{equation}
N^\mu({\bf x}) \equiv \int\! p^\mu f(t,{\bf x},{\bf p})\;\frac{{\rm d}^2{\bf p}}{E}.
\label{eq:Numu}
\end{equation}

In analytical calculations, implementing the above equations is straightforward.
The numerical implementation of $\alpha_2^{\bf p}$ in the transport algorithm is more challenging than in a hydrodynamic simulation, as we do not have the energy-momentum tensor directly given at each point in space and time.
To calculate the necessary quantities, we divided the system of test particles into $N_{\rm cells} = 40^2$ cells on a rectangular grid.%
\footnote{We have checked that our result is independent of the number of cells chosen in the computation by varying the number of cells from $N_{\rm cells} = 20^2$ to $N_{\rm cells} = 120^2$ and fixing $N_{\rm p} = 5\times 10^5$ particles.
  Fluctuations in the signal arise above $N_{\rm cells} = 60^2$, as the number of particles in the individual cells is too small.} 
We then computed the sum of the momentum components $p_x$ resp.\ $p_y$ of the particles in each cell, which we denote by ${\sf P}_x$ resp.\ ${\sf P}_y$.
For each cell, we also computed the total energy $E_{\rm tot.}$, the number of particles, and the mean velocity per particle $\bar{\sf v}_x\equiv {\sf P}_x/E_{\rm tot.}$ and $\bar{\sf v}_y\equiv {\sf P}_y/E_{\rm tot.}$.
Then we can approximate the numerator of Eq.~\eqref{eq:alpha2p} by
\begin{align}
\big\langle T^{0x}u_x-T^{0y}u_y \big\rangle_{\textbf{x}} \approx
\frac{1}{N_{\rm cells}}\sum_{i=1}^{N_{\rm cells}} 
  \frac{{\sf P}_{x,i} \bar{\sf v}_{x,i}-{\sf P}_{y,i} \bar{\sf v}_{y,i}}{\sqrt{1-\bar{\sf v}_{x,i}^2-\bar{\sf v}_{y,i}^2}}
\end{align}
and the denominator by
\begin{align}
\big\langle T^{00}u_0 \big\rangle_{\textbf{x}} \approx
\frac{1}{N_{\rm cells}}\sum_{i=1}^{N_{\rm cells}}
  \frac{E_{{\rm tot.},i}}{\sqrt{1-\bar{\sf v}_{x,i}^2-\bar{\sf v}_{y,i}^2}},
\end{align}
where the term in the denominators is due to the normalization of the velocity.

In the analytical approach starting from the Taylor expansion~\eqref{FullExpansion}, a necessary ingredient is the calculation in the free-streaming case. 
It turns out that in a collisionless system $\alpha_2^{\bf p}$ is not constant, but grows quadratically with time:
\begin{align}
\alpha_2^{\bf p}(t) \propto t^2 + \mathcal{O}(t^3),
\end{align}
where the proportionality factor of the term in $t^2$ involves the initial spatial eccentricity $\varepsilon_2^{\bf x}$.
This non-constant behavior is somewhat unexpected, since no anisotropic (here elliptic) flow in the usual acceptation of the term develops in a free streaming system. 

In turn, collisions modify this behavior at order $t^3$, consistent with the onset of $v_2(t)$ or $\varepsilon_2^{\bf p}(t)$, yet subleading compared to the collisionless evolution. 

\begin{figure}[!t]
	\centering
	\includegraphics[width=\columnwidth]{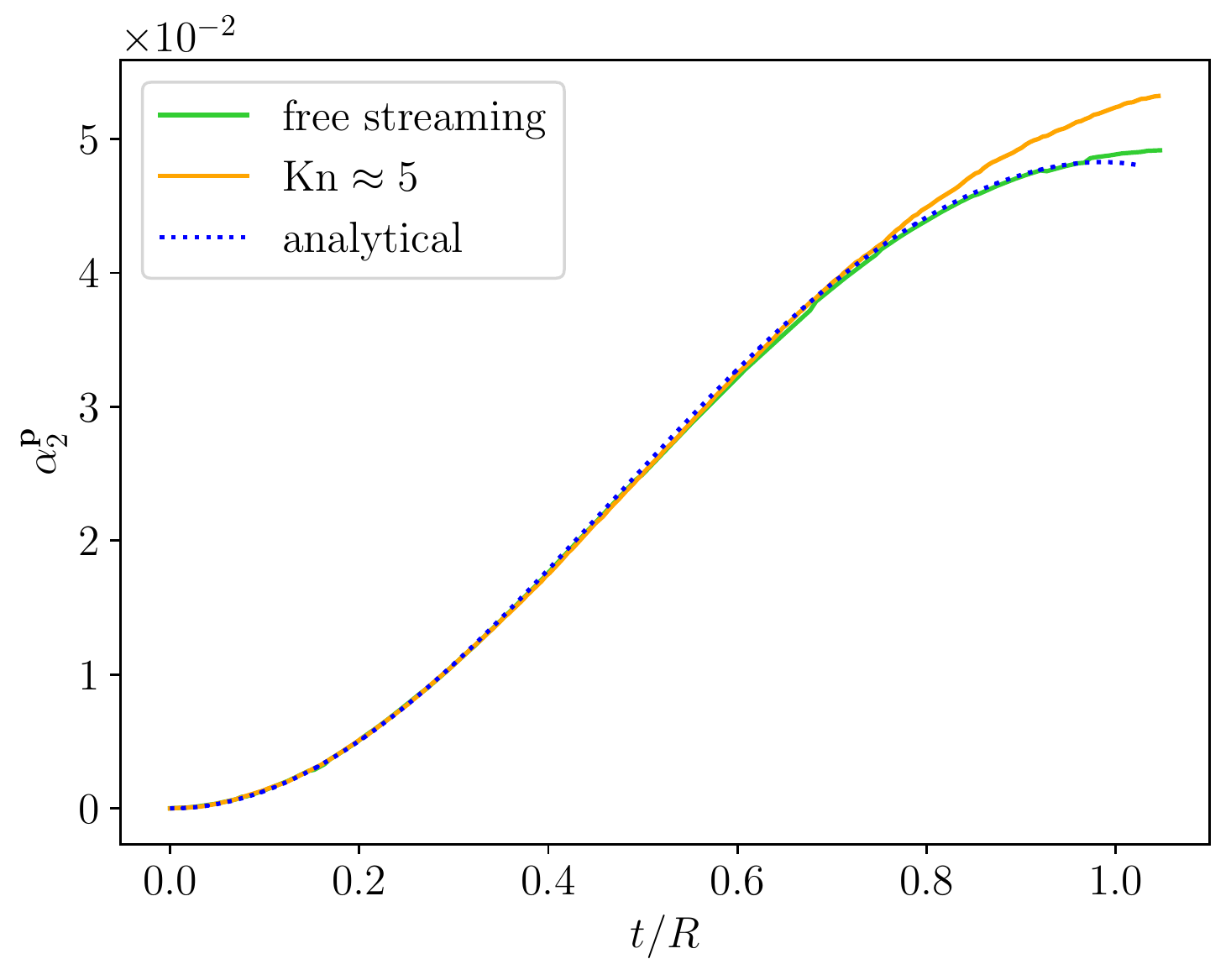}
	\caption{Time dependence of $\alpha_2^{\bf p}$ for simulations (full lines) in a collisionless (green) system or in the few collisions regime (orange), together with the analytical result in the free streaming case (dotted blue line).}
	\label{fig:alpha2p_fs}
\end{figure}

In Fig.~\ref{fig:alpha2p_fs} we show the early-time behavior of $\alpha_2^{\bf p}$ for a transport simulation with a moderate number of rescatterings (full orange line) with the $2\to 2$ collision kernel and in the free-streaming case (full green curve), together with the analytical calculation for the collisionless system (dotted blue curve). 
As anticipated, rescatterings only lead to a sizable departure from the free streaming behavior when $t/R\approx 0.8$, but their contribution is subleading at early times.

\section{Discussion}
\label{sec:discussion}

In this paper, we have studied the early time development of several quantities describing the asymmetry in the transverse plane, either in position or in momentum space, of a system of massless particles.
For that purpose, we used simulations from a numerical transport code and semi-analytical calculations based on a recently introduced~\cite{Borrell:2021cmh} Taylor expansion of the single-particle phase space density at early times.
Results from both approaches were compared in the few-rescatterings regime where the application of the analytical model makes sense. 
We also performed numerical simulations over a large range of Knudsen numbers, to study the early-time behavior of anisotropic flow coefficients from the few-rescatterings regime to the fluid-dynamical limit.

In a system with few rescatterings, we found that the Taylor-expansion based ansatz can indeed capture the numerical findings at early times, namely the scaling behavior $v_n(t) \propto t^{n+1}$ --- which also holds for the energy-weighted anisotropic flow coefficients $v_n^E(t)$ ---, the change in time of the number of rescatterings in the system, or the slow departure of geometrical characteristics like the spatial eccentricities $\varepsilon_n^{\bf x}(t)$ from their free-streaming evolution. 
We also showed that including higher-order terms in $t$ and/or in the cross section $\sigma$ improves the quality of the analytical predictions --- say typically until $t/R\approx 0.5$. 
However, we should quickly add that any new order in $t$ or in $\sigma$ typically means the appearance of many more terms in the calculations. 
It would thus be highly desirable to find resummation schemes if one wishes to push the analytical expansion to larger times.\footnote{For $v_2(t)$ in the $2\to 0$ scenario and with our specific case of initial condition, one can find an analytical expression resumming all order in $t$ at ${\cal O}(\sigma)$~\cite{Bachmann:2022cls}, and it seems feasible also at ${\cal O}(\sigma^2)$, although we did not attempt it.}

Letting the number of rescatterings per particle grow, so that we can also reproduce the hydrodynamic limit, we found that the early-time scaling behavior of $v_n(t)$ changes smoothly from $t^{n+1}$ at large Knudsen number to $t^n$ when Kn is small. 
We quantified this change via a non-integer scaling exponent $v_n(t)\propto t^{\gamma_n}$, which we cannot find directly in the Taylor-series approach. 
It could be that a double resummation in $t$ and $\sigma$ can indeed provide such a non-analytic behavior, but it could also be an artifact from our fitting ansatz and procedure.
In any case, we view $\gamma_n$ as an effective exponent characterizing the behavior of $v_n(t)$ on times of order 0.1--0.5$R$.
On such a time scale set by the transverse system size, the fluid-dynamical scaling behavior appears as the many-collision limit of kinetic theory. 
This convergence in studies of anisotropic flow was found previously (for the $v_n$ values at the end of the system evolution) in Refs.~\cite{Gombeaud:2007ub,Drescher:2007cd}, which use a similar two-dimensional setup as ours, or Ref.~\cite{Ambrus:2021fej} in the case of a longitudinally boost-invariant expansion.\footnote{There the authors emphasize that their kinetic transport calculations in the relaxation time approximation (RTA) only converge towards the hydrodynamic behavior if they consider the large-opacity limit at fixed initialization time $\tau_0$ of the system.}

Instead of defining early times in comparison to $R$ as we did, one could replace the latter by the mean free path $\ell_\mathrm{mfp} = {\rm Kn}\cdot R$.
In that case, we would expect that for times smaller than say about $0.3\ell_\mathrm{mfp}$, kinetic theory yields the behavior $v_n(t)\propto t^{n+1}$ of the few-collisions regime~\cite{Borrell:2021cmh}, even for values of Kn ``in the fluid-dynamical limit'', that is much smaller than 1.
Unfortunately, investigating this ``very early'' behavior in a setup with ${\rm Kn}\ll 1$ is unfeasible with our cascade code if we wish to remain in the dilute regime. 
Yet with this alternative definition of ``early times'', depending --- at fixed initial geometry --- on the interaction strength, we would probably find different scaling laws for the flow coefficients according as the system is described by kinetic theory (irrespective of the Knudsen number) or fluid dynamics. 
This would be similar to the finding that kinetic theory in the RTA and hydrodynamics have different early-time attractors for the ratio of longitudinal pressure over energy density~\cite{Kurkela:2019set}. 

While we cannot claim that our findings are of immediate relevance for heavy-ion phenomenology, although this is not excluded for the case of small systems, yet we think that they may still be of interest, in particular for studies of the initial stages~\cite{Schlichting:2019abc}.
Whatever dynamical model is used for the prehydrodynamic evolution, some amount of anisotropic flow will develop --- unless of course it is assumed that the system is freely streaming ---, and it is not uninteresting to know whether the model is ``hydro-like'' or rather ``few-collisions-like'', or somewhere inbetween.
One could even assume as initial-stage model a set of simple scaling laws like $v_n(t)\propto t^{\gamma_n}$ and similar for other properties of the system, and see how this affects global Bayesian reconstructions of the properties of the medium created in heavy ion collisions.

More on the theoretical side, our results are relevant for searches for the existence of dynamical attractor solutions for systems with transverse dynamics~\cite{Ambrus:2021sjg}: even ``late time attractors'' have to accommodate qualitatively different ``early time'' behaviors. 
In that respect, it would be interesting to extend our study to spatially asymmetric setups for which analytical results in the hydrodynamic limit are known~\cite{Csanad:2014dpa}.

\begin{acknowledgments}
  We thank Nina Feld, Sören Schlichting and Clemens Werthmann for discussions, as well as M\'at\'e Csan\'ad and Tam\'as Cs\"org\H o for helpful comments on a preliminary version of our results. 
  The authors acknowledge support by the Deutsche Forschungsgemeinschaft (DFG, German Research Foundation) through the CRC-TR 211 'Strong-interaction matter under extreme conditions' - project number 315477589 - TRR 211.
  Numerical simulations presented in this work were performed at the Paderborn Center for Parallel Computing (PC$^2$) and the Bielefeld GPU Cluster, and we gratefully acknowledge their support.
\end{acknowledgments}

\appendix

\section{Energy-weighted mean square radius}
\label{appendix:<Er^2>}

In this Appendix, we show that in a two-dimensional system of massless particles described by the Boltzmann equation --- with a ``physical'' collision kernel ${\cal C}[f]$ implementing energy and momentum conservation ---, the evolution of the energy-weighted mean square radius of the system actually does not depend on the presence of collisions, i.e.\ it behaves as if the system were collisionless. 

Mathematically, the conservation of energy and momentum in the collisions translates into the identity
\begin{equation}
\label{C[f]_integral}
\int\!p^\mu {\cal C}[f]\,\frac{{\rm d}^2\bf p}{E} = 0,
\end{equation}
at any position ${\bf x}$ and for any $\mu\in\{0,1,2\}$, where $p^0\equiv E$ denotes the energy. 
Differentiating this equation with respect to time and exchanging time derivative and integration over momentum space, one deduces for any $j\geq 0$
\begin{equation}
\label{d_tC[f]_integral}
\int\!p^\mu \partial_t^j {\cal C}[f]\,\frac{{\rm d}^2\bf p}{E} = 0,
\end{equation}
again valid at any ${\bf x}$ and where the time derivative can be evaluated at any time, in particular in the initial state. 

Now let us look at the early-time Taylor expansion~\eqref{FullExpansion} of the phase space distribution $f(t,{\bf x},{\bf p})$. 
Multiplying it by the energy $E = |{\bf p}|$, one finds that the terms due to collisions at order $t^k$ with $k\geq 1$ are all of the form
\[
\bigg(\!\!-\!\frac{{\bf p} \cdot \bm{\nabla}_{\!x}}{E}\bigg)^{\!\!j}
   \partial_t^{k-1-j}{\cal C}[f]\big|_0.
\]
When calculating the energy-weighted mean square radius, i.e.\ the denominator of Eq.~\eqref{eccentricity_E} with $n=2$, we thus encounter at order $t^k$ integrals of the form
\begin{equation}
\label{<Er^2>_generic-term}
\int\!r^2 \bigg(\frac{{\bf p} \cdot \bm{\nabla}_{\!x}}{E}\bigg)^{\!\!l}
\partial_t^{k-1-l}{\cal C}[f]\big|_0\,{\rm d}^2{\bf x}\,{\rm d}^2{\bf p}
\end{equation}
with $k\geq 1$ and $0\leq l\leq k-1$. 
We shall several times use the property that since the system has a finite size, $f$ and in turn ${\cal C}[f]$ and all its derivatives vanish at (spatial) infinity.

When $l=0$ --- as happens for instance for the linear term in $t$ in the Taylor expansion ---, expression~\eqref{<Er^2>_generic-term} takes the form of the integral over ${\bf x}$ of $r^2$ times an integral of the type~\eqref{d_tC[f]_integral} with $j=k-1$ and $\mu = 0$. 
Thus all those terms vanish thanks to energy conservation.

In the case $l=1$, we first consider the integral over ${\bf x}$ in Eq.~\eqref{<Er^2>_generic-term}, and integrate once by parts to get rid of the gradient. 
This replaces for instance $x^2 p_x\partial_x$ by $-2x p_x$ in the integrand, while the integrated term, involving $\partial_t^{k-2}{\cal C}[f]|_0$ at infinity, is zero.
The remaining integral is of the form 
\[
-2\!\int\!\frac{xp_x + yp_y}{E}_{}
  \partial_t^{k-2}{\cal C}[f]\big|_0\,{\rm d}^2{\bf x}\,{\rm d}^2{\bf p}. 
\]
Fixing ${\bf x}$ and looking at the integral over momentum space, it is of the type~\eqref{d_tC[f]_integral} with $\mu=1$ or 2, and thus equals 0 due to momentum conservation.

For $l=2$, a twofold integration by parts over ${\bf x}$ replaces $r^2({\bf p} \cdot \bm{\nabla}_{\!x})^2\partial_t^{k-3}{\cal C}[f]|_0$ in the integrand of Eq.~\eqref{<Er^2>_generic-term} by $2(p_x^2+p_y^2)_{}\partial_t^{k-3}{\cal C}[f]|_0$. 
Since the particles are massless and propagate in 2 dimensions, $p_x^2+p_y^2$ equals $E^2$, which cancels out the denominator of the integrand. 
The integral thus becomes
\[
2\!\int\!\partial_t^{k-3}{\cal C}[f]\big|_0\,{\rm d}^2{\bf x}\,{\rm d}^2{\bf p},
\]
which is zero due to energy conservation [Eq.~\eqref{d_tC[f]_integral} for $\mu=0$].

When $l\geq 3$, one can again transform the term~\eqref{<Er^2>_generic-term} by integrating twice by parts over ${\bf x}$. 
Discarding the vanishing integrated terms, the effect of the twofold integration by parts amounts to the symbolical replacement of $r^2 (\bm{\nabla}_{\!x})^l$ by $2(\bm{\nabla}_{\!x})^{l-2}$ in the integrand.
Since $l-2\geq 1$, what remains in the integrand is still a spatial derivative of a function that vanishes as $|{\bf x}|\to\infty$: a further integration over space thus yields 0. 

All in all, we thus find that all integrals of the form~\eqref{<Er^2>_generic-term} vanish for the specific case under study. 
This is not the case for massive particles or for particles with $p_z\neq 0$.

Anticipating on the following Appendix, note that we used twice --- for the cases $l=2$ and $l\geq 3$ --- a twofold integration by parts over ${\bf x}$ to get rid of $r^2$ and two spatial derivatives in the integrand. 
If $r^2$ is replaced by $r^3$, then the reasoning used to find that the terms with $l=2$ vanish no longer works.

\section{Early time evolution of the numerator and denominator of the triangularity $\varepsilon_3^{\bf x}$}
\label{appendix:epsilon3}

\begin{figure}[!t]
    \centering
    \includegraphics[width=\columnwidth]{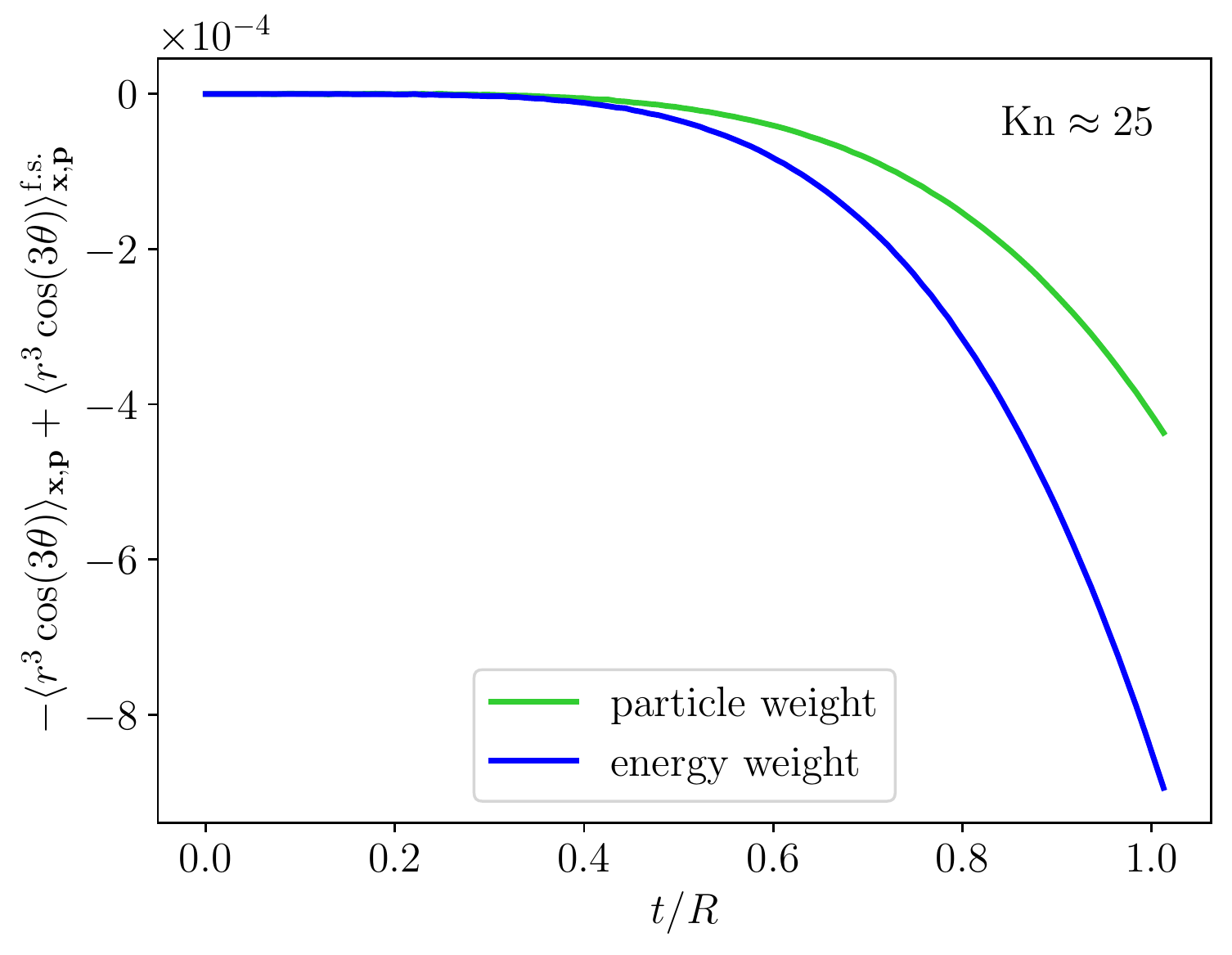}
    \includegraphics[width=\columnwidth]{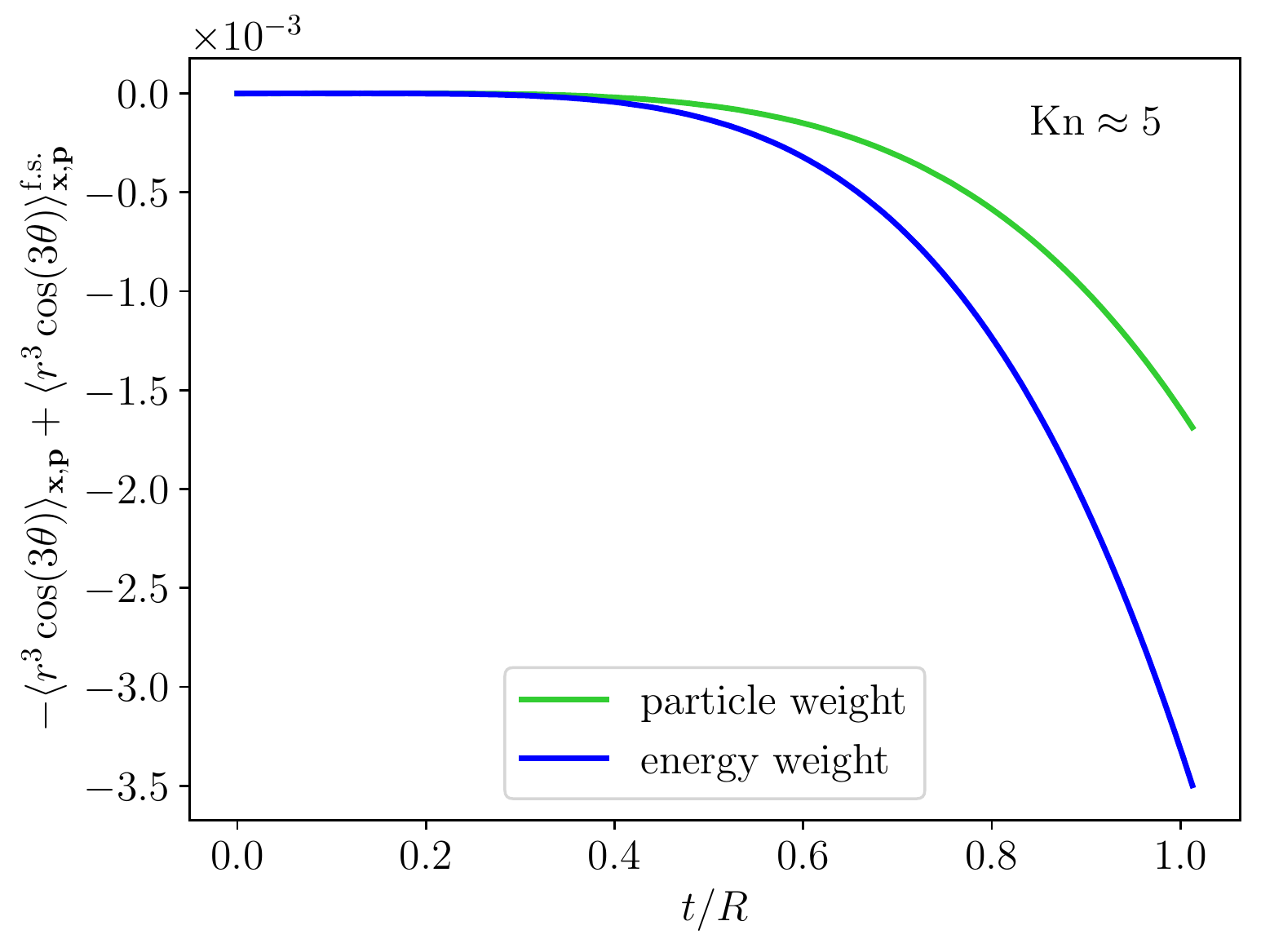}
    \caption{Time evolution of the departure of $-\langle r^3\cos(3\theta)\rangle_{{\bf x},{\bf p}}$ with particle-number (green) or energy (blue) weighting from its free streaming value, for simulations with $\mathrm{Kn}\approx 25$ (top panel) or $\mathrm{Kn}\approx 5$ (bottom panel).}
    \label{fig:rcubedcos3theta}
\end{figure}
\begin{figure}[!t]
	\centering
	\includegraphics[width=\columnwidth]{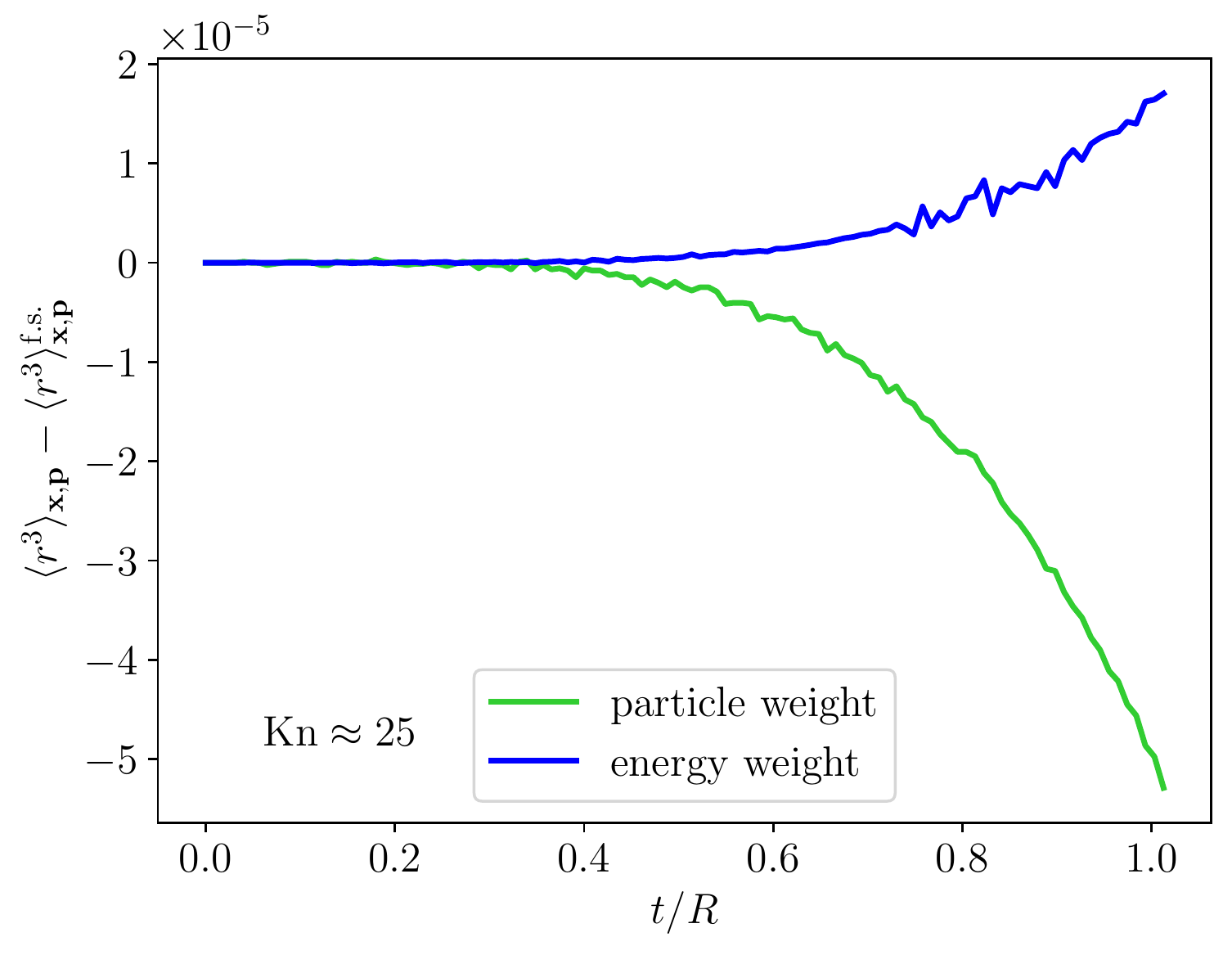}
	\includegraphics[width=\columnwidth]{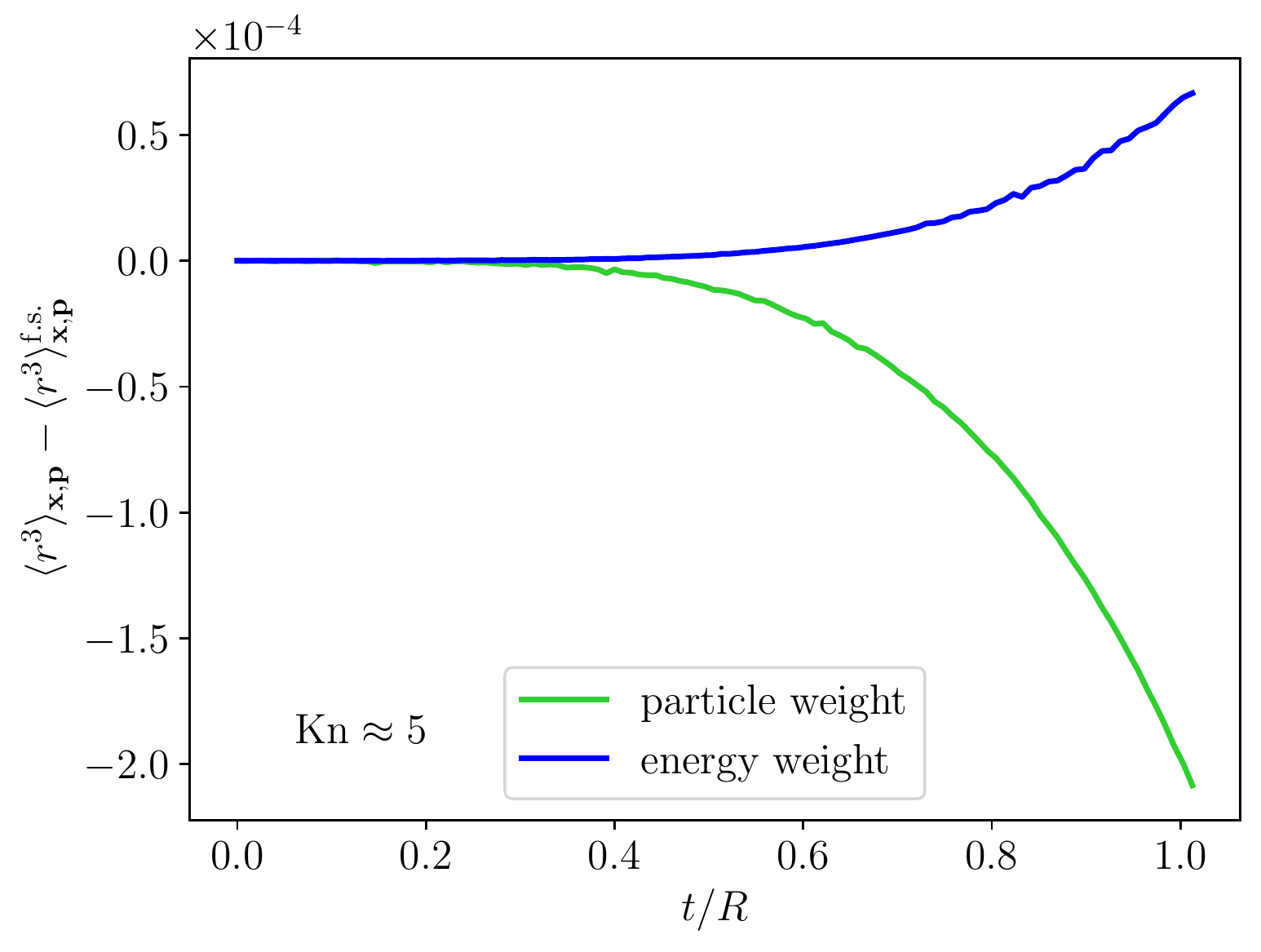}
	\caption{Time evolution of the departure of $\langle r^3\rangle_{{\bf x},{\bf p}}$ with particle-number (green) or energy (blue) weighting from its free streaming behavior, for simulations with $\mathrm{Kn}\approx 25$ (top panel) or $\mathrm{Kn}\approx 5$ (bottom panel).}
	\label{fig:rcubed}
\end{figure}

In this Appendix we discuss the influence of rescatterings on the numerator and denominator of the spatial triangularity $\varepsilon_3^{\bf x}$, namely the particle-number or energy weighted average values of $-r^3\cos(3\theta)$ and $r^3$.
Their departure from the corresponding free streaming behaviors, scaled by their respective values in the initial state, are shown in Figs.~\ref{fig:rcubedcos3theta} and \ref{fig:rcubed} for simulations at ${\rm Kn} = 25$ (top panels) and ${\rm Kn} = 5$ (bottom panels). 

A first observation is that these relative deviations from the collisionless case roughly scale linearly with the number of rescatterings, since they are about 5 times larger at ${\rm Kn} = 5$ than at ${\rm Kn} = 25$. 
This is similar to what was obtained in the second harmonic in Sect.~\ref{subsec:eccentricities}.

Fitting the early time behaviors with single power laws, like those of Eqs.~\eqref{eq:r2cos2thetafit} and \eqref{eq:r2fit}, gave us rather inconclusive values of the scaling exponents in the range 4--5.7. 
Again, given the smallness of the signal and the slowness of its evolution, we do not know how much we can trust those results, as our early-time fits may be dominated by numerical noise (which is clearly visible in both panels of Fig.~\ref{fig:rcubed}).

A difference with the second-harmonic results of Sect.~\ref{subsec:eccentricities} is that the energy-weighted mean cubed radius deviates from its free streaming behavior. 
And indeed, in our analytical calculations we find that the arguments that allowed us to show that $\langle r^2\rangle_{{\bf x},{\bf p}}$ with energy weight is zero do not hold for $\langle r^3\rangle_{{\bf x},{\bf p}}$.
Interestingly, one sees that the rate of growth of $\langle r^3\rangle_{{\bf x},{\bf p}}$ with energy weight is increased by rescatterings, while that of $\langle r^3\rangle_{{\bf x},{\bf p}}$ with particle-number weight is decreased.
This means that rescatterings redistribute the energy density in the system, and more precisely they tend to transport energy from the inner regions --- which had the largest energy density in the initial state --- towards the outer region, which seems quite intuitive.

\section{Elliptic momentum anisotropy $\varepsilon_2^{\bf p}$ at large times}
\label{app_eps2p}

In this Appendix, we present for completeness a few results from our calculations of the momentum space eccentricity $\varepsilon_2^{\bf p}$ that go beyond the ``early-time behavior'' that is the main scope of the paper.

\begin{figure*}[!t]
	\centering
	\includegraphics[width=\columnwidth]{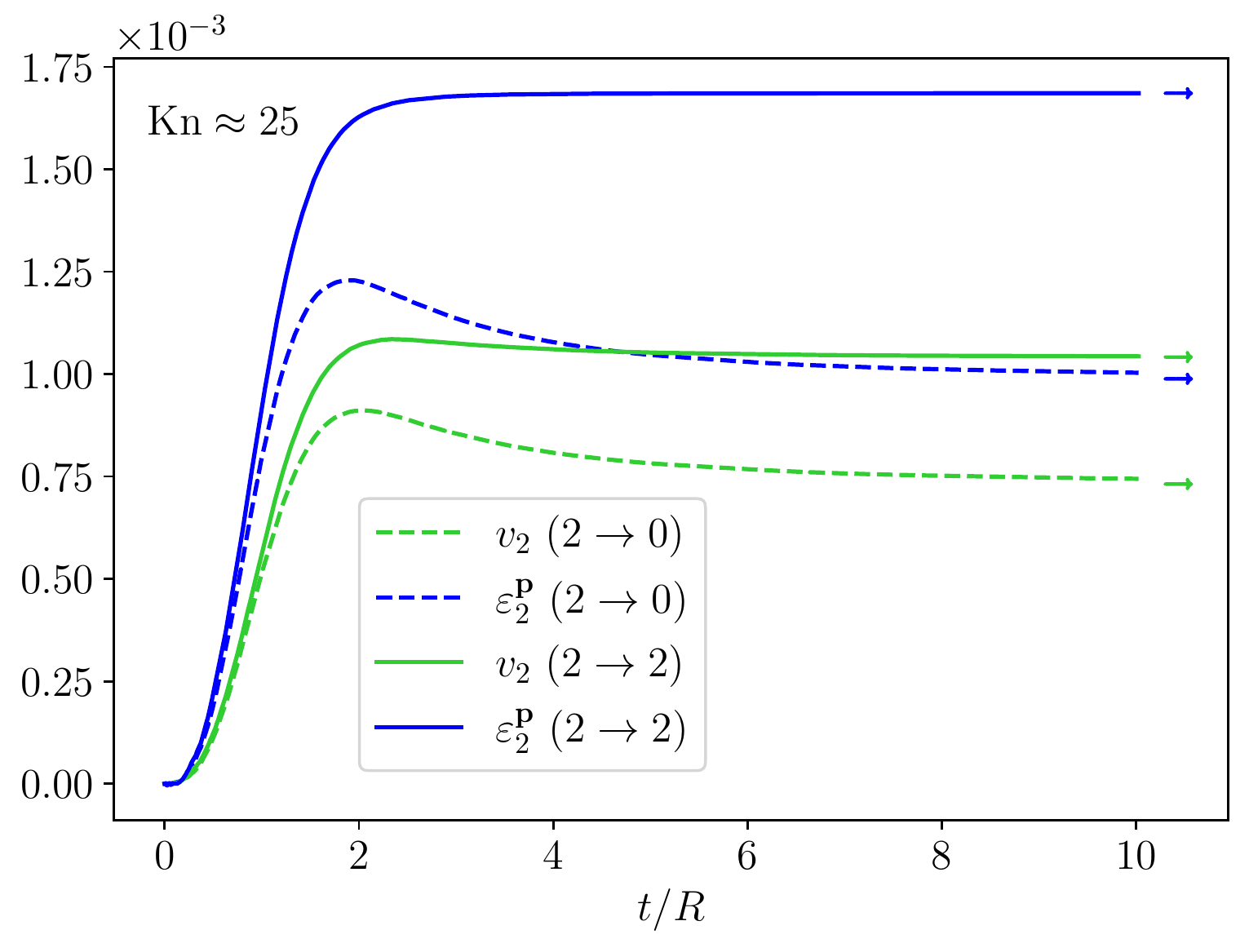}
	\includegraphics[width=\columnwidth]{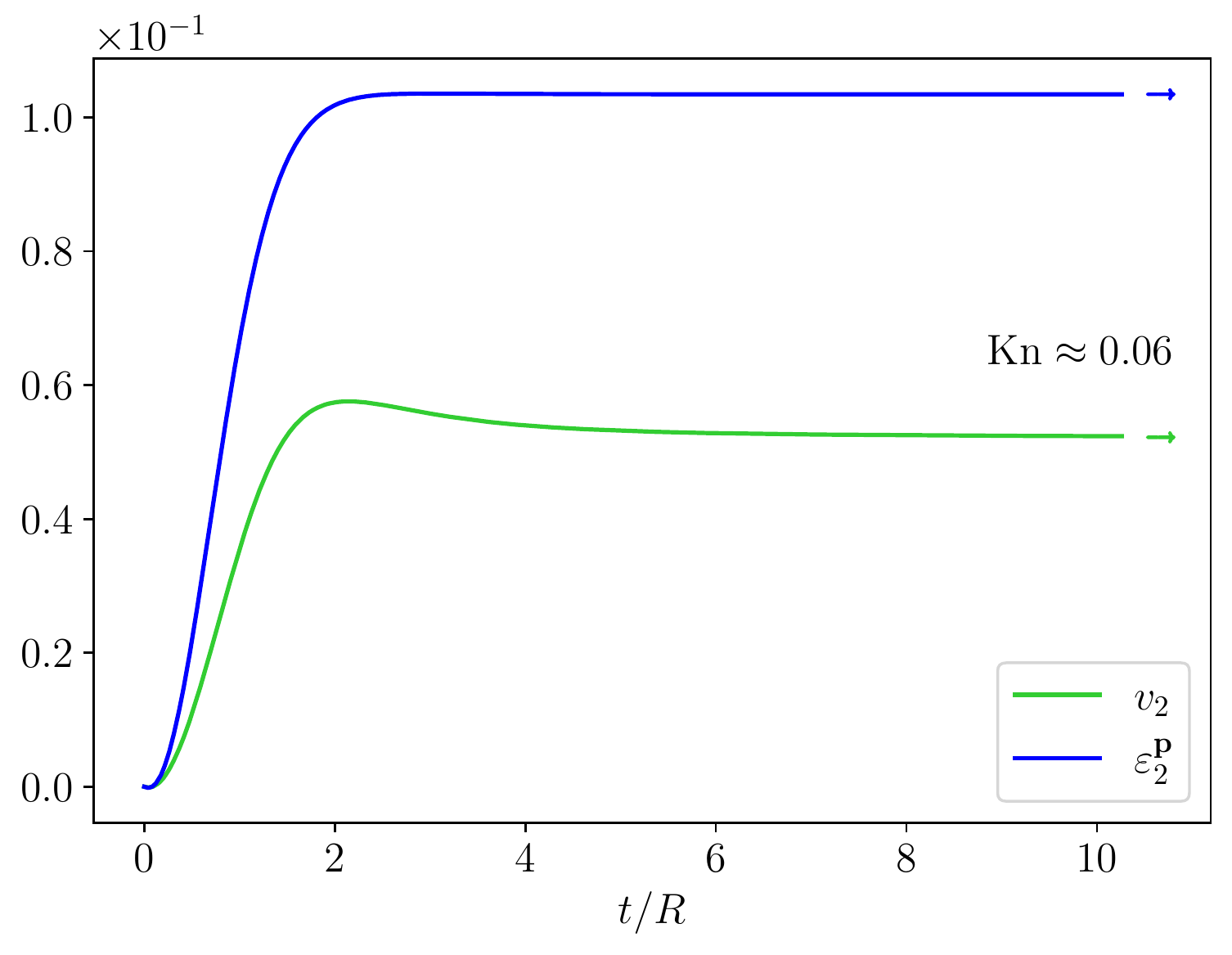}
	\caption{Evolution of the anisotropic flow coefficient $v_2$ (green) and the eccentricity $\varepsilon_2^{\bf p}$ in momentum space (blue) for the $2\to 2$ collision kernel in the few collisions limit (left) and in the hydrodynamic limit (right). 
		The arrows indicate the corresponding values at $t/R\approx 30$.}
	\label{fig:v2_eps2_late_time}
\end{figure*}

Figure~\ref{fig:v2_eps2_late_time} compares the evolution of $\varepsilon_2^{\bf p}$ --- which we recall coincides with the energy-weighted elliptic flow $v_2^E$ in our setup --- and the particle-number weighted $v_2$. 
This is shown both for calculations with elastic binary collisions (full curves), in which case both the few-rescatterings regime (left panel: ${\rm Kn}\approx 25$) and the fluid-dynamical limit (right panel: ${\rm Kn}\approx 0.06$) are illustrated, and in the $2\to 0$ scenario (dashed curves), in which only the calculations at large Knudsen number make sense. 
All simulations shown in Fig.~\ref{fig:v2_eps2_late_time} were performed with the same initial phase space distribution, with $\varepsilon_2^{\bf x}\simeq 0.15$.

Qualitatively, $\varepsilon_2^{\bf p}(t)$ and $v_2(t)$ behave similarly: 
both rise until $t/R \approx 2$, and then they either saturate or slightly decrease ($2\to 2$ collision kernel). 
In the case of $v_2(t)$ this agrees well with what was found using a slightly different initial profile but a similar transport algorithm~\cite{Alver:2010dn}.
The decrease is more pronounced (about 20\%) in the $2\to 0$ scenario, which is discussed more extensively in Ref.~\cite{Bachmann:2022cls}.

On a quantitative level, $\varepsilon_2^{\bf p}$ is systematically larger than $v_2$, irrespective of the model under consideration. 
In the fluid dynamical regime, one finds $\varepsilon_2^{\bf p} \approx 2v_2$ at late times, as already argued in the literature~\cite{Bhalerao:2005mm} for two-dimensional expanding systems.
Yet overall, $v_2$ and its energy-weighted version $\varepsilon_2^{\bf p} = v_2^E$ have parallel behaviors, both in the $2\to 2$ or $2\to 0$ scenarios. 
This contrasts with their triangular counterparts $v_3$ and $v_3^E$, which as we saw in Sect.~\ref{subsec:flow_coefficients} and \ref{subsec:alternative_flow_observables} behave differently in the $2\to 0$ model.

\end{document}